\documentclass[12pt, draftclsnofoot, onecolumn]{IEEEtran}
\IEEEoverridecommandlockouts
\usepackage{graphicx,amssymb,amsmath,color}
\usepackage{epstopdf}
\usepackage[mathscr]{euscript}
\usepackage{dsfont}

\usepackage[noadjust]{cite}
\usepackage[font=footnotesize]{caption}
\usepackage[font=footnotesize]{subcaption}
\usepackage[labelsep=period]{caption}

\usepackage{bm} 
\usepackage{multirow} 
\usepackage{psfrag}
\usepackage{array,booktabs}
\usepackage{txfonts}
\usepackage{lipsum}

\usepackage{enumerate}
\usepackage{enumitem}

\usepackage[english]{babel}
\usepackage{blindtext}
\usepackage{mathtools}

\usepackage{relsize}
\usepackage[subnum]{cases}

\usepackage[flushleft]{threeparttable}

\usepackage[hyphens]{url}

\usepackage[flushleft]{threeparttable}
\usepackage{adjustbox}
\usepackage{tabularx}

\addto\captionsenglish{\renewcommand{\figurename}{Fig.}}

\newcommand{\nc}{\newcommand}

\newcommand\figref{Fig.~\ref}
\newcommand\figsref{Figs.~\ref}

\nc{\widthoffigs}{1}

\nc{\ith}{$i^{\text{th}}$ }
\nc{\jth}{$j^{\text{th}}$ }
\nc{\nth}{$n^{\text{th}}$ }

\nc{\ale}{\alpha_{\overrightarrow{e_0}}}
\nc{\aln}{\alpha_{\overrightarrow{n_0}}}
\nc{\bee}{\beta_{\overrightarrow{e_0}}}
\nc{\ben}{\beta_{\overrightarrow{n_0}}}

\nc{\FoV}{\mathrm{FoV}}
\nc{\onePsi}{\mathds{1}(\Psi<\mathrm{FoV})}
\nc{\cPsi}{\cos(\Psi)}

\nc{\SIR}{\text{SIR}}
\nc{\ex}{\text{e}}
\nc{\taumin}{\tau_{\text{min}}}
\nc{\taumax}{\tau_{\text{max}}}

\nc{\iagg}{i_\text{agg}}
\nc{\Iagg}{I_\text{agg}}

\nc{\lb}{\left(}
\nc{\lsb}{\left[}
\nc{\rb}{\right)}
\nc{\rsb}{\right]}

\nc{\BDC}{B_{\text{DC}}}

\newcommand*{\QEDB}{\hfill\ensuremath{\blacksquare}}%

\newcommand{\HX}[1]{H_{\text{{#1}}}}
\newcommand{\GX}[1]{G_{\text{{#1}}}}

\newcommand{\EX}[1]{{\mathbb{E}}\left[{#1}\right]}

\newcommand{\PX}[1]{{\mathbb{P}}\left[{#1}\right]}

\newcommand{\ccos}[1]{\cos\left( {#1} \right)}
\newcommand{\ssin}[1]{\sin\left( {#1} \right)}

\newcommand{\colbluehh}[1]{{\color{black}{#1}}}
\newcommand{\colredmoh}[1]{{\color{black}{#1}}}

\newcommand{\colredms}[1]{{\color{black}{#1}}}
\newcommand{\colbluemss}[1]{{\color{black}{#1}}}
\newcommand{\colredmss}[1]{{\color{black}{#1}}}
\newcommand{\colredmsss}[1]{{\color{black}{#1}}}

\nc{\degr}{^{\circ}}

\begin{document}
%
\title{Terminal Orientation in OFDM-based LiFi Systems}


\author{Ardimas~Andi~Purwita,~\IEEEmembership{Student Member,~IEEE,}
        Mohammad~Dehghani~Soltani,~\IEEEmembership{Student Member,~IEEE,}
        Majid~Safari,~\IEEEmembership{Member,~IEEE,}
        and~Harald~Haas,~\IEEEmembership{Fellow,~IEEE}
 \thanks{The authors are with School of Engineering, Institute for Digital Communications, LiFi R\&D Centre, The University of Edinburgh, Edinburgh EH9 3JL, U.K. (e-mail:~\{a.purwita, m.dehghani, majid.safari, h.haas\}@ed.ac.uk).}
}

\maketitle

%
\begin{abstract}
\boldmath Light-fidelity (LiFi) is a wireless communication technology that employs both infrared and visible light spectra to support multiuser access and user mobility. Considering the small wavelength of light, the optical channel is affected by the random orientation of a user equipment (UE). In this paper, a random process model for changes in the UE orientation is proposed based on data measurements. We show that the coherence time of the random orientation is in the order of hundreds of milliseconds. Therefore, an indoor optical wireless channel can be treated as a slowly-varying channel as its delay spread is typically in the order of nanoseconds. A study of the orientation model on the performance of direct-current-biased orthogonal frequency-division multiplexing (DC-OFDM) is also presented. The performance analysis of the DC-OFDM system incorporates the effect of diffuse link due to reflection and blockage by the user. The results show that the diffuse link and the blockage have significant effects, especially if the UE is located relatively far away from an access point (AP). It is shown that the effect is notable if the horizontal distance between the UE and the AP is greater than $1.5$ m in a typical $5\times3.5\times3$ m$^3$ indoor room.
\end{abstract}
    
\begin{IEEEkeywords}
LiFi, measurement, random orientation, random process model, OFDM.
\end{IEEEkeywords}

\section{Introduction}
    \IEEEPARstart{L}{}ight-fidelity (LiFi) has recently attracted significant research interest as the scarcity in the radio frequency (RF) spectrum becomes a major concern, see \cite{fccspectrum} and \cite{lifihaas}. Based on predictions of mobile data traffic, the number of cell sites and achievable spectral efficiency, the entire RF spectrum in the US will be fully utilized in around the year 2035 \cite{tezcan5g}; hence, \colredms{the} use of optical wireless communications is \colbluemss{of great interest}. The benefit of LiFi over other optical wireless communication systems, such as visible light communication (VLC), is that LiFi supports \colbluemss{multi-user} connections and bi-directionality. In other words, LiFi provides similar functionality as the other \colbluehh{Institute of Electrical and Electronics Engineers (IEEE)} 802.11 technologies, e.g., handover and multiple access. \colbluehh{Therefore, a working group for LiFi in IEEE 802.11 already exists \cite{lifiwg} as well as a task group for LiFi in IEEE 802.15 \cite{lifitig}.}


    In an indoor room, the quality of the optical channel highly depends on the geometry parameters of a user equipment (UE), an access point (AP) and the dimensions of the room \cite{infraredkahnbarry}. Therefore, a significant change in the parameters due to the mobility of the users greatly affects the channel. The main focus of this paper is the random orientation of the UE. Meanwhile, the majority of studies on LiFi or VLC only consider the case where the UE faces upward, see \cite{lifihaas} or \cite{downlinkcheng} and references therein. The facing-upward assumption is practically \colbluehh{supported by LiFi universal serial bus (USB) dongles used by a laptop or a tablet computer. LiFi USB dongles have been commercialized, and this use case is already deployed.} The next \colbluehh{natural step} for LiFi is \colbluehh{for it to be integrated into} mobile devices, such as smartphones \colbluehh{and indeed} tablets or laptops. In practice, the orientation of the mobile devices is random in nature. Therefore, it is of importance to study the characteristics of the random orientation.

    Over the past few years, \colbluehh{only a few papers have been published which have considered the random orientation of the mobile terminal}, see \cite{aveneauorientation,orientationsoltani,wangtilted,wangber,erogluorientation,yapicinoma,ardimaswcnc,MDSHandover,ardimashandover,soltaniaccess}. The authors in \cite{aveneauorientation,wangber} and \cite{erogluorientation} focus on the bit error ratio (BER) performances of an on-off keying (OOK) system with a random receiver orientation. Handover probability in a LiFi network with randomly-oriented UEs is studied in \cite{orientationsoltani} and \cite{ardimashandover}. The studies of the VLC channel capacity and the effect of random orientation to non-orthogonal multiple access (NOMA) are presented in \cite{wangtilted} and \cite{yapicinoma}, respectively. All the authors assume that the random orientation follows a certain probability density function (PDF) without any support from data measurements. \colredms{In our initial work \cite{ardimaswcnc,soltanitcom}}, the PDF of the random orientation based on \colbluehh{a} series of experiments is presented. \colbluemss{A generalized random way point model is also proposed in \cite{soltanitcom} considering \colredmsss{the} effect of orientation using an autoregressive (AR) model.} Another experiment result is reported in \cite{orientationpeng}; however, the authors \colbluemss{only focus on} the \colbluehh{rate of change of orientation}. In this paper, a more comprehensive analysis of the experimental results \colbluemss{first} reported in \cite{ardimaswcnc} is provided.

    \colredms{The proposed PDFs in \cite{ardimaswcnc} and \cite{soltanitcom}} \colbluemss{model the random orientation as a random variable} by directly counting the frequency of data, which are assumed to be \colbluemss{uncorrelated.} In other words, temporal characteristics of the random orientation, e.g., the autocorrelation function (ACF), are not described. Moreover, the noise from the measurement device is ignored. In this paper, \colbluemss{the random orientation is characterized based on a random process (RP) model.} The main challenge is that the time sampling of data measurements is not evenly-spaced. Therefore, the ACF cannot be directly estimated from the sample data, and the conventional Fourier analysis cannot be used since the sinusoids are no longer orthogonal. Therefore, the least-squares spectral analysis (LSSA) \cite{vaniceklssa,scargle1982,craymerdissertation,baisch1999,vanderplas} is used to estimate the spectral characteristics of the data measurements. In addition, a Wiener filter is used to filter the noise in the data measurement \cite{hayesbook}, and the CLEAN\footnote{To the best of authors' knowledge, it is not an abbreviation and first introduced as CLEAN algorithm.} algorithm is applied to eliminate the partial-aliasing in the power spectrum, see \cite{baisch1999} and \cite{robertsclean}. 

    \colredms{Based on our proposed RP model, it is shown that the channel is highly-correlated within the order of \colbluemss{hundreds of milliseconds}.} The RP model is suitable for the type of studies which consider the movement of the users inside a room, such as the \colbluehh{studies} of handover probability in a LiFi network \cite{orientationsoltani} and \cite{ardimashandover}. \colredms{In addition, \colbluemss{the RP model can be used in mobility models such as the orientation-based random waypoint model proposed in} \cite{soltanitcom}.} In this paper, the random orientation model is applied to an orthogonal frequency-division multiplexing (OFDM)-based LiFi system \cite{lifihaas}. The main advantage of OFDM over other typical modulation systems, such as OOK, is that it can mitigate the inter-symbol interference (ISI) which is caused by multipath propagation of the indoor optical channel, see \cite{infraredkahnbarry} and \cite{barrykahnsim}.
    
    

    As a use case, this paper focuses on the comparison of the analysis of the LiFi system with and without the diffuse link considering the random orientation model. The reason \colredms{for this} is that most studies in LiFi or VLC, see \cite{aveneauorientation,orientationsoltani,wangtilted,wangber,erogluorientation,yapicinoma,ardimaswcnc} and \cite{ardimashandover}, neglect the diffuse link, and only \colredmoh{the line-of-sight (LOS)} channel is considered \colredms{due to the fact that} the received signal power is dominated by the power from the LOS link. It will be shown later that this is not always the case, and it will significantly affect the \colbluehh{performance} of our LiFi \colbluemss{system in some scenarios}. This result is consistent with other results in \cite{infraredkahnbarry,barrykahnsim} and \cite{highorderreflection} which consider the diffuse link. However, these studies only focus on the OOK modulation scheme. In this paper, the effect of the diffuse link in the OFDM-based LiFi system is investigated. We will also consider a human body as \colredmss{both blocking and reflector objects} since the random orientation is modeled based on data measurements that were collected while the participants were using the UEs.



    In Section II, our system model is presented. Our experimental setups and results are discussed in Section III and IV, respectively. The use cases of the random orientation model to the OFDM-based LiFi system are presented in Section V and VI. In Section V, the performances of the LiFi system with fixed locations of the UEs and the users are studied. The results in Section V are then generalized to the case where the locations and the orientations of the UEs are random in Section VI. Section VII concludes our work.

\section{System Model}
    
\begin{figure}
            \centering
            \begin{subfigure}[b]{.5\columnwidth}
                \centering
                \includegraphics[width=1\columnwidth,draft=false]{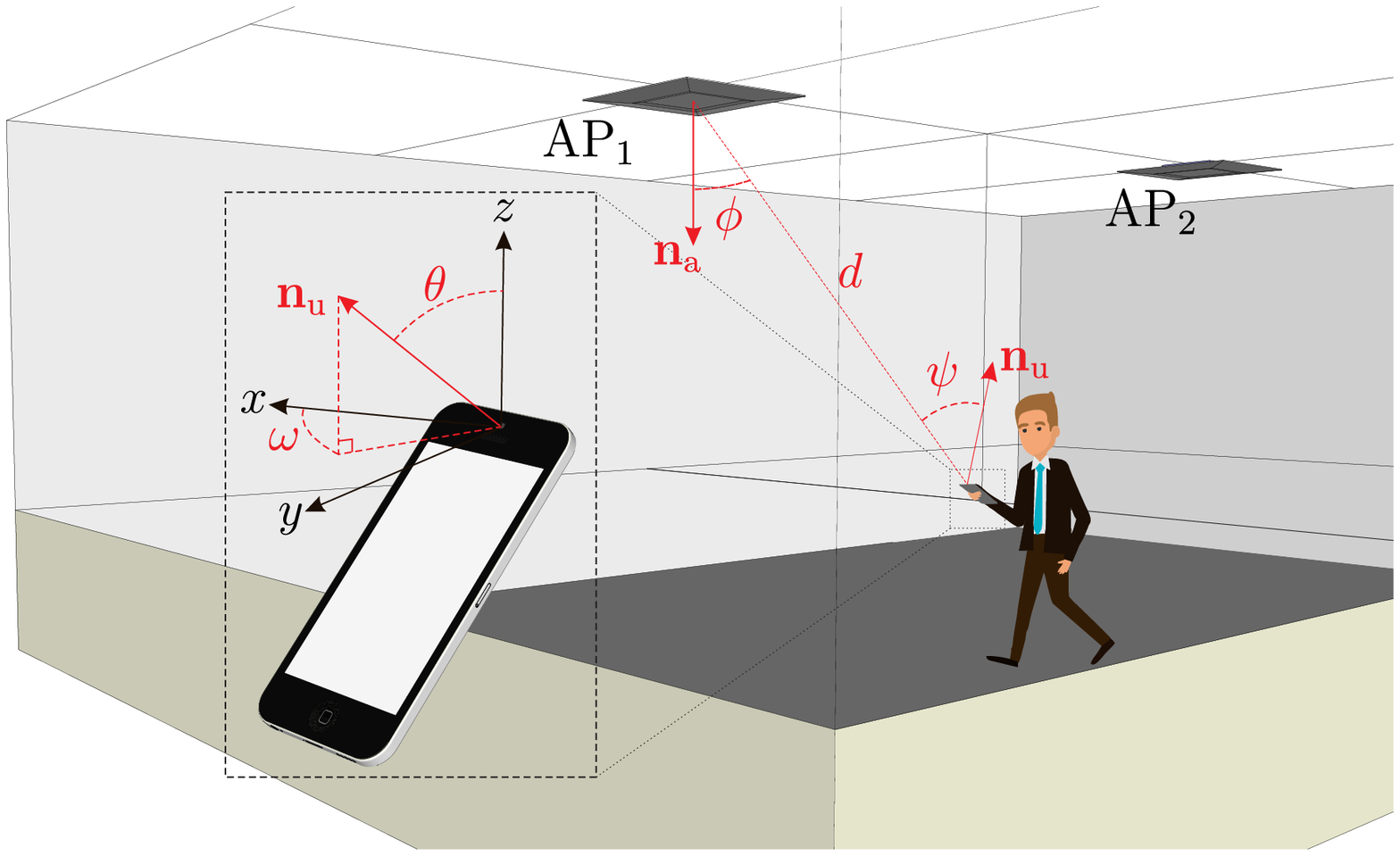}
                \caption{}
            \end{subfigure}~
            \begin{subfigure}[b]{.5\columnwidth}
                \centering
                \includegraphics[width=1\columnwidth,draft=false]{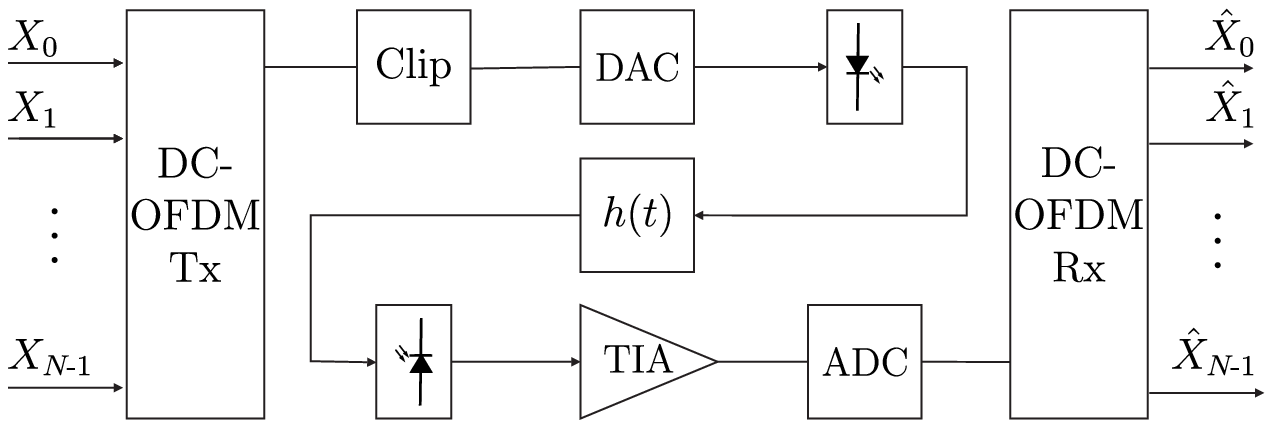}
                \vspace{.4cm}
                \caption{}
            \end{subfigure}
            \caption{(a) System model and (b) block diagram of a DC-OFDM-based system.}
            \label{figsystemmodel}
        \end{figure}

    An indoor room with several optical APs, known as the optical attocell \cite{lifihaas}, and a mobile user using his LiFi-enabled device, referred to as a UE, is assumed as depicted in \figref{figsystemmodel}(a). Throughout this paper, the AP acts as a transmitter (Tx), and the UE acts as a receiver (Rx). A UE is assumed to have a photodiode (PD) \colredms{on the front screen} and held in a comfortable manner, which will be discussed in the next section. A direct-current-biased OFDM (DC-OFDM) based system is used in this paper; the block diagram of the system is shown in \figref{figsystemmodel}(b). In this section, we first explain our OFDM transmission system and then the channel model. 



    \subsection{Transmission System}

        In the Tx, $N$ \colredmoh{inverse discrete Fourier transform IDFT} operations are performed to obtain discrete time and parallel OFDM symbols, which are expressed by:
            \begin{align*}\label{}
                x_{\text{p}}[k] = \sum\limits_{n=0}^{N-1} X_n \text{e}^{ j 2 \pi n k/N }, \quad k = 0, 1, \dots, N-1,
            \end{align*} 
        \noindent where $X_n$ is the symbol that is transmitted at the \nth subcarrier \cite{perinofdm}. The symbol $X_n$ is chosen from an $M_n$-ary quadrature amplitude modulation (QAM) constellation with the average power $\EX{\left| X_n \right|^2} = P_n$. Moreover, $X_n$ should follow the Hermitian symmetry, i.e., $X_n = X_{N-n}^*$. In this paper, the constellation size and the power per symbol at each subcarrier are fixed, i.e., $M_n = M$ and $P_n = 1$.

        If the number of subcarriers, $N$, is large, it is widely known that $x_{\text{p}}[k]$ follows a Gaussian distribution with zero mean and variance, $\sigma^2 = 2 \left(\frac{N}{2}-1 \right)$ \cite{perinofdm}.\footnote{Recall that the DC subcarrier is used for the DC bias.} Furthermore, for a given bit rate $R_\text{b}$, the OFDM symbol rate is $R_\text{s} = 2 R_\text{b} / \log M$.\footnote{The factor $2$ is due to the Hermitian symmetry.} After adding the cyclic prefix (CP), the OFDM sampling rate is:
            \begin{align} \label{}
                f_\text{s} = r_{\text{os}} R_\text{s} \frac{N+N_{\text{cp}}}{N},
            \end{align}
        \noindent where $N_{\text{cp}}$ is the length of CP, and $r_{\text{os}}$ is the oversampling ratio, i.e., $r_{\text{os}} = N/N_\text{u}$, where $N_\text{u}$ is the number of subcarriers that are used. Zero padding in the frequency domain is applied for the unused subcarriers. The bandwidth of the OFDM symbol is defined as $f_\text{s}/(2 r_{\text{os}})$.

        Let's now define $x[k]$ as a discrete time OFDM signal after parallel-to-serial conversion and adding the CP. The signal $x[k]$ needs to be clipped because (i) to meet the dynamic range of the digital-to-analog converter (DAC) and the analog-to-digital converter (ADC), (ii) to meet the non-negative constraint of the intensity modulation (IM) and (iii) to avoid the nonlinearities of the LED. Let $x_\text{c}[k]$ be a discrete time clipped signal of $x[k]$ at levels $-r_1 \sigma$ and $r_2 \sigma$\colredmoh{, where $r_1$ and $r_2$ denote the clipping ratios}. It is assumed that a predistortion is employed such the \textit{P-I} curve of the LED in the non-clipped region is linear \cite{predistortionelgala}. Using the Gaussian approximation, i.e., $x[k]$ follows the Gaussian distribution, the Bussgang's Theorem can be applied; therefore, $x_\text{c}[k]$ can be written as $x_\text{c}[k] = K x[k] + d[k]$, where $d[k]$ is a random process that is uncorrelated with $x[k]$, and $K$ is defined as $K = 1-\left( Q(r_1) + Q(r_2) \right)$, \colredmoh{where $Q(\cdot)$ is the tail distribution function of a normal distribution \cite{perinofdm}.} In other words, $K$ shows the attenuation level due to the clipping process. 

        The zero-order hold is used to model the DAC. Let $\HX{DAC}(f)$ be a frequency response of it with the cutoff frequency is equal to the bandwidth of the OFDM symbols. Note that the low-pass frequency in the zero-order hold is modeled by a fifth order Bessel filter \cite{perinofdm}. A DC bias is performed after the DAC to meet the non-negative constraint and defined as $ 2 r \sum_{n = 1}^{\frac{N}{2}-1} \lvert \HX{DAC}(f_n) \rvert^2$, where $f_n$ denotes the frequency of the \nth subcarrier. The frequency response of the LED is denoted by $\HX{LED}(f)$ and modeled as a low pass filter using a first order Butterworth filter with cutoff frequency $f_{\text{c}_{\text{LED}}}$. The output signal of the LED, which is in the optical domain, is transmitted over a channel with a channel impulse response (CIR) $h(t)$ whose frequency response is $\HX{CIR}(f)$. 
        

        In the receiver, the frequency responses of the PD and the transimpedance amplifier (TIA) are represented by the frequency response of an antialiasing filter in the ADC, $\HX{ADC}(f)$, as it is typically narrower than the others. Note that the PD has a responsitivy denoted by $R$ whose physical unit is Ampere per Watt (A/W). As in the DAC, the cutoff frequency of ADC is chosen to be equal to the bandwidth of the OFDM symbols. We mainly vary the cutoff frequency of the LED, $f_{\text{c}_{\text{LED}}}$, as it is typically the main limiting factor in the front-ends of an optical wireless communication system \cite{downlinkcheng}. In this paper, only the thermal noise is considered, and its single-sided power spectral density (PSD) is denoted by $N_0$. A \colredms{single} tap equalizer is used to estimate the transmitted symbols.

        In general, the BER at the \nth subcarrier, $P_{\text{b}_n}$, can be approximated as follows \cite{perinofdm}:
            \begin{align}\label{eqberpersubcarrier}
                P_{\text{b}_n} \approx \frac{4}{\log_2 M}\left(1 - \frac{1}{\sqrt{M}}\right)Q \left(\sqrt{\frac{3 \gamma_n}{M-1}}\ \right),
            \end{align}
        \noindent where $\gamma_n$ is the electrical signal-to-noise ratio (SNR) at the \nth subcarrier, which can be expressed as follows:
            \begin{align}\label{eqsnrpersubcarrier}        
                \gamma_n = \frac{K^2 R^2 \lvert \HX{DAC}(f_n) \HX{LED}(f_n) \HX{CIR}(f_n) \HX{ADC}(f_n) \rvert^2}{f_\text{s} N_0 \lvert \HX{ADC} \rvert^2/\left( 2 N \right)}.
            \end{align}
        \noindent The average BER for all OFDM symbols is denoted by $P_\text{b}$ which is simply the average of $P_{\text{b}_n}$ over all subcarriers.

    \subsection{Channel Impulse Response, $h(t)$}

        The multipath propagation in an indoor optical wireless channel is described by the CIR $h(t)$ and its frequency response $\HX{CIR}(f)$. A widely used method to calculate the CIR is proposed by Kahn and Barry in \cite{barrykahnsim}. This method is significantly improved by Schulze in \cite{schulzefreqdomain} by taking into account all reflections. This method is used in this paper since our interest is not to observe how adding an increased number of high order reflections affects the LiFi analyses, but instead to observe how neglecting the reflections affect the analyses.





        In this paper, an empty office room with dimensions length ($L$) $\times$ width ($W$) $\times$ height ($H$) m$^3$ is assumed. A human body will be modeled as a rectangular prism, which is similar to \cite{carruthersiterative}, i.e., the iterative version of Kahn and Barry's method. Considering an object as a rectangular prism using Schulze's method is straightforward, and it will be discussed later. A similar approach is presented in \cite{laveneaubody}, but the authors also model other common body parts. From our observations, adding such details is less significant to our analyses than focusing on the reflectivities of those body parts. In addition, diffractions on the edge of a human body are ignored in this paper since the wavelengths of the infrared and visible light spectrum are relatively short compared to the dimension of the edge of a human body. Blockages due to other people are not considered in this paper. 
        

        The CIR $h(t)$ can be decomposed into the LOS and the diffuse links \cite{schulzefreqdomain}. \colbluemss{That is, it can be written as:}
            \begin{align}\label{eqhtHf}
                h(t) = h_\text{Rx,Tx}(t) + h_{\text{diff}}(t) \xLeftrightarrow[\mathcal{F}^{-1}]{\mathcal{F}} \HX{CIR}(f) = \HX{Rx,Tx}(f) + \HX{diff}(f),
            \end{align}
        \noindent where $\mathcal{F}$ denotes the Fourrier transform. Let's define an attenuation factor between the Rx and the Tx as follows:
            \begin{align}\label{eqGRxTx}
                \GX{Rx,Tx} = \frac{m+1}{2 \pi} \cos^m \left(\phi_{\text{Rx,Tx}}\right) \frac{A_{\text{Rx}} \cos \left(\psi_{\text{Rx,Tx}}\right) }{d^2_{\text{Rx,Tx}}} v_{\text{Rx,Tx}}(\FoV),
            \end{align}
        \noindent where $m$ is the Lambertian index, $A_{\text{Rx}}$ denotes the receiver area and $\psi_{\text{Rx,Tx}}$ is the incident angle or angle of arrival between the normal vector $\mathbf{n}_\text{u}$ and the position of the Rx. Note that the orientations of an LED in the Tx and a PD in the RX are shown by the unit normal vector $\mathbf{n}_\text{u}$ and $\mathbf{n}_\text{a}$, respectively. The radiant angle or angle of departure between the normal vector $\mathbf{n}_\text{a}$ and the position of the Tx is denoted by $\phi_{\text{Rx,Tx}}$. The distance between the Tx and Rx is denoted by $d_{\text{Rx,Tx}}$. The term $v_{\text{Rx,Tx}}$ has a binary value that denotes a visibility factor of the link between an Rx and a Tx, i.e., $v_{\text{Rx,Tx}}$ is one if $0 \leq \phi_{\text{Rx,Tx}} \leq \pi/2$ and there is no blocking object between the link Tx and Rx. The visibility factor is also a function of the field-of-view (FoV) of a receiver. That is, $v_{\text{Rx,Tx}}(\FoV)$ is one if $0 \leq \psi_{\text{Rx,Tx}} \leq \FoV$. 

        The LOS link, $\HX{Rx,Tx}(f)$, can be expressed by \cite{schulzefreqdomain}:
            \begin{align}\label{eqHRxTx}
                \HX{Rx,Tx}(f) = \GX{Rx,Tx} \exp\left( -j 2\pi f d_{\text{Rx,Tx}}/c \right),
            \end{align}
        \noindent where $c$ is the speed of light. The diffuse component can be calculated by partitioning each surface into discrete smaller elements, i.e., $p = 1, 2, \dots, P$, and treating them as another radiator or receiver. These discrete elements include the partition of the surfaces of the blocking objects as they are modeled by rectangular prisms. This is the reason why the calculation of $h(t)$ when a human body is modeled by a rectangular prism is straightforward. The smaller elements act as a radiator with the Lambertian index $m = 1$. The gain of the reflected light has an additional attenuation factor, which is called the reflectivity, $\rho$.

        The frequency response of the diffuse component can be calculated as follows \cite{schulzefreqdomain}:
            \begin{align}\label{eqHdiff}
                \HX{diff}(f) = \mathbf{r}^{\text{T}}(f) \mathbf{G}_\rho \left(\mathbf{I} - \mathbf{G}(f)\mathbf{G}_\rho \right)^{-1} \mathbf{t}(f).
            \end{align}
        \noindent The vector $\mathbf{r}^{\text{T}}(f) = \left( \HX{RX,1}(f)\ \HX{RX,2}(f)\ \cdots\ \HX{RX,$P$}(f) \right)$ is called the receiver transfer vector. The transmitter transfer vector is denoted by $\mathbf{t}(f) = \left( \HX{1,TX}(f)\ \HX{2,TX}(f)\ \cdots\ \HX{$P$,TX}(f) \right)^{\text{T}}$. The room intrinsic frequency response for each small element is described by a $P \times P$ matrix $\mathbf{G}(f)$ whose its \ith row and \jth column represent the LOS frequency response between the \ith and the \jth elements. The matrix $\mathbf{G}_\rho$ denotes the reflectivity matrix where $\mathbf{G}_\rho = \text{diag}\left(\rho_1, \rho_2, \dots, \rho_P \right)$.
\section{Experiment}
    
    \subsection{Assumptions}
        The unit normal vector of the transceiver is denoted by the vector $\mathbf{n}_{\text{u}}$, and spherical coordinates are used to describe its orientation. A polar angle is denoted by $\theta$, and an azimuth angle is denoted by $\omega$, see the inset in \figref{figsystemmodel}(a). It is clear that from (\ref{eqhtHf}-\ref{eqHdiff}), as the normal vector $\mathbf{n}_\text{u}$ changes, so does the CIR. Therefore, it is important to have the random orientation model to study the CIR's behavior. In this section, our experimental results on the random orientation of a UE are discussed.

\begin{figure}
                \centering
                \begin{subfigure}[b]{.5\columnwidth}
                    \centering
                    \includegraphics[width=1\columnwidth,draft=false]{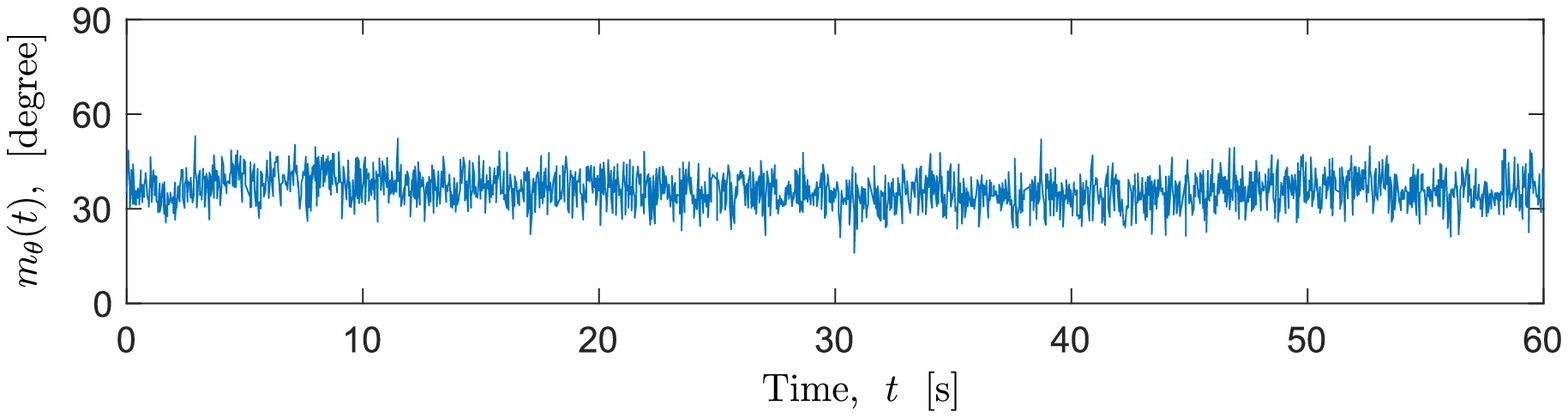}
                    \includegraphics[width=1\columnwidth,draft=false]{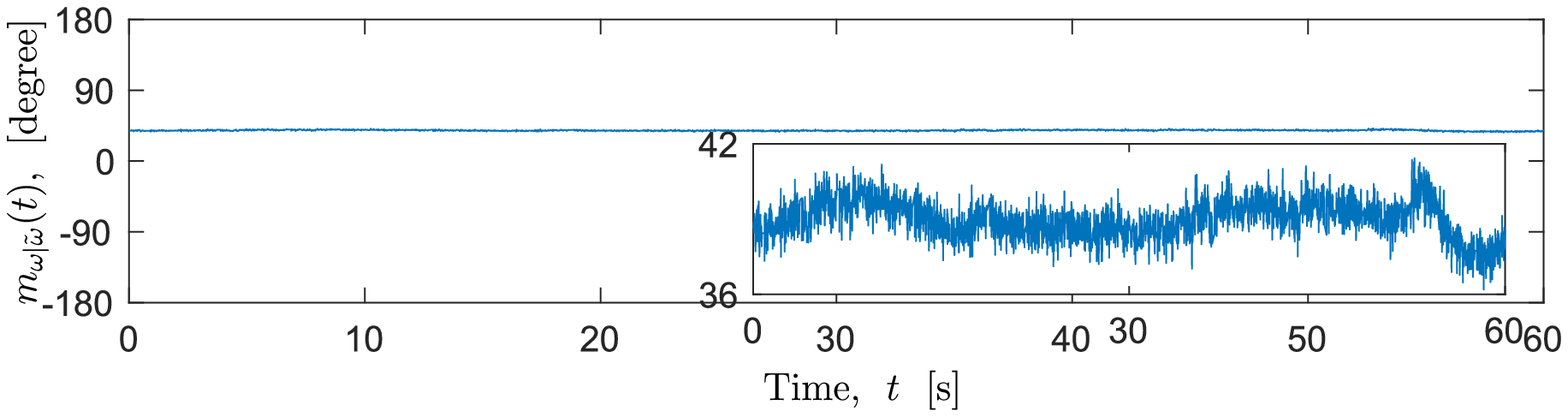}
                    \caption{}
                \end{subfigure}~
                \begin{subfigure}[b]{.5\columnwidth}
                    \centering
                    \includegraphics[width=1\columnwidth,draft=false]{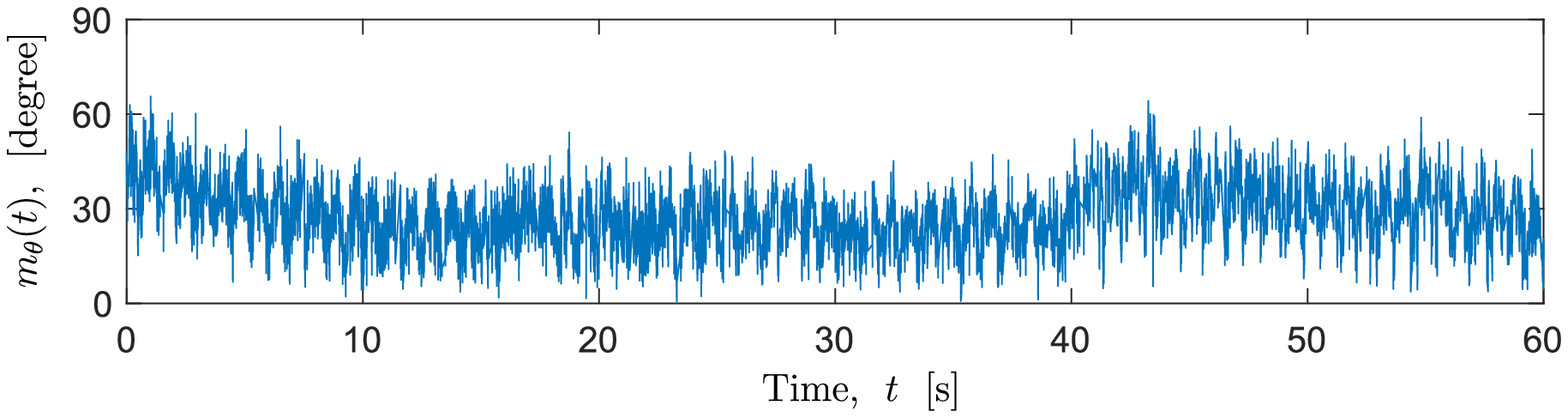}
                    \includegraphics[width=1\columnwidth,draft=false]{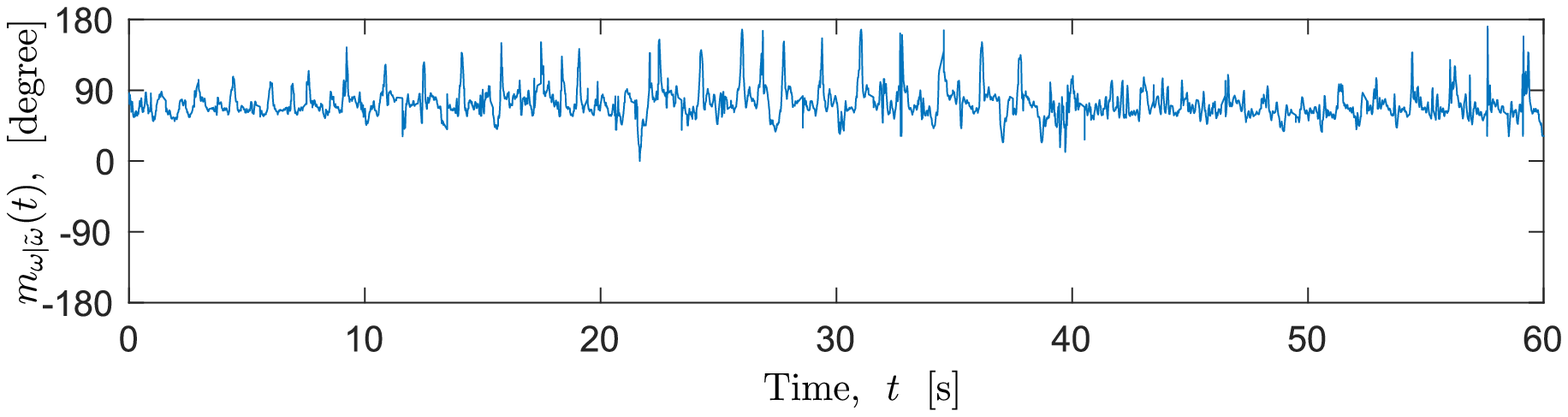}
                    \caption{}
                \end{subfigure}
                \caption{Samples of noisy measurement: (a) $m_\theta(t)$ and $m_{\omega|\tilde{\omega}}(t)$ for the sitting activity and (b) $m_\theta(t)$ and $m_{\omega|\tilde{\omega}}(t)$ for the walking activity.}
                \label{figsamples}
            \end{figure}

        \colredmss{From our experimental data, it is observed that the noisy measurement of $\theta$, which is denoted by $m_\theta(t)$, fluctuates \colredmsss{around} its mean as depicted in \figref{figsamples}. Note that discussions about our experimental setups are given in the next section. In this paper, the random orientation is modeled as the change of orientation that is caused by \colbluemss{user behavior such as} hand movements and other activities, such as typing or scrolling. \colbluemss{Therefore, we can write it as:}
            \begin{align}\label{eqdefrp}
                m_\theta(t) &= \mu_\theta + \theta_0(t) + n_\theta(t), 
            \end{align}
        \noindent where $\mu_\theta$ denotes the mean of $m_\theta(t)$, the subscript `$0$' refers to the corresponding RP whose mean is zero and $n_\theta(t)$ denotes the noise of the measurement. Furthermore, we assume that both $\theta_0(t)$ and $n_\theta(t)$ are wide-sense stationary (WSS) and independent to each other. \colredmsss{While the} WSS RP asserts the mean to be constant, it will be shown later that our proposed model gives a reasonable match with the experimental data in terms of the ACF and the power spectrum.
        
        For $\omega$, it needs a different assumption as it \textit{highly} depends on the directions of the users. That is, for a different sample, $\omega$ might no longer fluctuate around $40^\circ$ as in \figref{figsamples}(b), and it could fluctuate around $-40^\circ$ if the participant faced the other way. This does not happen in the $\theta$ case as all measurement data $\theta$ generally fluctuate around $\mu_\theta$. For the azimuth angle $\omega$, \colredmsss{conditioned on the direction of the UE, $\tilde{\omega}$,} we use following noisy measurement model:
        \begin{align}\label{eqdefrpomega}
            m_{\omega|\tilde{\omega}}(t) = \tilde{\omega} + \omega_0(t) + n_\omega(t),
        \end{align}
        \noindent where $\tilde{\omega}$ is the angle direction of the UE and will be modeled as a RV \colredmsss{following the uniform distribution over the interval $(-\pi,\pi]$ as shown in \cite{soltanitcom} and \cite{zhihongorientation}}. Therefore, the subscript on $\tilde{\omega}$ denotes that \colredmsss{$m_{\omega|\tilde{\omega}}(t)$} is valid for a given value of $\tilde{\omega}$. The physical meaning of $\tilde{\omega}$ is the direction of a UE, which will later be used to define the direction of a user who holds it. The descriptions of $\omega_0(t)$ and $n_\omega(t)$ are the same as in \eqref{eqdefrp}.}
        

    \subsection{Experimental Setup}
        \colbluemss{The same experimental setup has been presented in \cite{ardimaswcnc} and \cite{soltanitcom}. It will be provided here for the convenience of the reader.} We asked $40$ participants from the Alexander Graham Building, University of Edinburgh to conduct a series of experiments. Nine of them were left-handed, and none suffered from a severe hand tremor. After being briefed on the procedure and the purpose of the experiments, they were asked to browse the Internet and watch videos as they streamed for a minute. Both activities are chosen to emulate the typical data that is obtained during activities requiring Internet services. To support our WSS assumption, the time limit was set to be one minute since the participants might get tired and change the means drastically if the experiment was longer. To be able to conduct the series of the experiments while measuring the data, the Physics Toolbox Sensor Suite application is used \cite{androidapp}. The application is installed on two different Samsung Galaxy S5 smartphones.


        The participants were also asked to sit, which is referred to as the \textit{sitting activity}, and walk, which is referred to as the \textit{walking activity}, during the experiments. The walking activity was conducted in a straight corridor with dimensions $40$ m $\times\ 15$ m. There was a total of $222$ measurements that were collected, which comprise $148$ data measurements for the sitting activity and $74$ data measurements for the walking activity.  
    
\section{Experimental Results and Analysis}

    The Physics Toolbox Sensor Suite application captures real values of rotations of a UE in terms of pitch, roll and yaw rotations. These values are then transformed into the spherical coordinates \cite{orientationsoltani}. It is important to note here that time sampling of the application is not evenly spaced and random in each data measurement. The \colredmss{time sampling duration} can range from $1$ ms to $0.65$ s. In fact, the most frequent \colredmss{time sampling durations} from all data measurements are, in the order, $1$ ms, $18$ ms and $64$ ms, for a minute experiment, each data measurement has around $2,000$ to $3,000$ samples. Instead of the conventional Fourier analysis, the LSSA that is based on the Lomb-Scargle method is used in this paper \cite{scargle1982}. 
    
    It should be noted here that several preprocessing steps are performed as the analysis is very sensitive to the time sampling. First, it is observed that some time samplings are not unique. That is, from the measurement data, there are some time samplings that are the same, and they give slightly different values. In this case, one sampling time is randomly chosen among them. Second, we noticed that a few participants already got tired keeping their arms up. Hence, we did not process the measurement data for that case since we are interested in the scenario where the mean of a measurement data is relatively constant (or we are interested in the slight hand movement about the mean). None of these steps are performed in \cite{ardimaswcnc,soltanitcom} or \cite{zhihongorientation} as the authors only focus on occurrence of the orientation sampling.  

    \figref{figsamples} shows some samples of data measurements that were collected. Notice that the samples fluctuate around their means. The mean $\mu_\theta$, is calculated as follows:
        \begin{align}\label{eqmeansample}
            \mu_\theta = \frac{1}{N}\sum\limits_{i = 1}^{N} \hat{\mu}_{\theta}^{(i)},
        \end{align}
    \noindent where $\hat{\mu}_{\theta}^{(i)}$ is the estimated mean of the \ith data measurement. {By taking average of all data measurements, we obtain ${\mu_\theta = 41.13\degr}$ for the sitting activity and ${\mu_\theta = 27.75\degr}$ for the walking activity.} These results show that when the participants sit, they tilt the phone to the back more than they would when they walk. The reason for this is because most of the participants put their elbows on a desk while using their phones. For the rest of paper, a comfortable manner is defined as the position where the user holds the UE with $\theta$ in the vicinity of $\mu_\theta$.

    \colredmss{Unlike $\mu_\theta$ in \eqref{eqdefrp}, if the user holds the UE in a normal position as depicted in \figref{figsystemmodel}(a), then $\tilde{\omega}$ in \eqref{eqdefrpomega} depends on which direction the user sits or walks.} \colredmsss{As also shown in \cite{soltanitcom} and also through an uncontrolled measurement in \cite{zhihongorientation}, the PDF of the unconditional azimuth angle $\omega$ is dominated by the random direction of the UE (i.e., $\tilde{\omega}$) and it can be modelled as uniformly distributed in $(-\pi, \pi]$. Therefore, the measurement average for azimuth angle does not provide a physical intuition.}
    


    \subsection{Noise Measurements}
    The characteristics of the data measurements with no activity are first observed; in other words, the noise of the application is measured. In particular, the spectrum and the behavior of the noise will be investigated. For the random sampling, it is suggested by \cite{scargle1982,baisch1999} and \cite{vanderplas} that we need to be concerned about the spectrum of the random sampling. Let $w(t)$ be a window function of the random sampling as expressed by:
        \begin{align}\label{}
            w(t) = 
                \begin{cases}
                1 ,& t = t_k \\
                0 ,& t \neq t_k. 
                \end{cases}
        \end{align}
    \noindent where $\{t_k\}_{k = 0}^{N-1}$ is the random sampling. It is important because there might be pseudo-aliasing or partial-aliasing of signals (see Figs. 13 and 14 in \cite{vanderplas} for clear examples). For these purposes, we collected $15$ data measurements containing noise samples with different time intervals from the same smartphones used in the experiments.

\begin{figure}
            \centering
            \begin{subfigure}[b]{.5\columnwidth}
                \centering
                \includegraphics[width=1\columnwidth,draft=false]{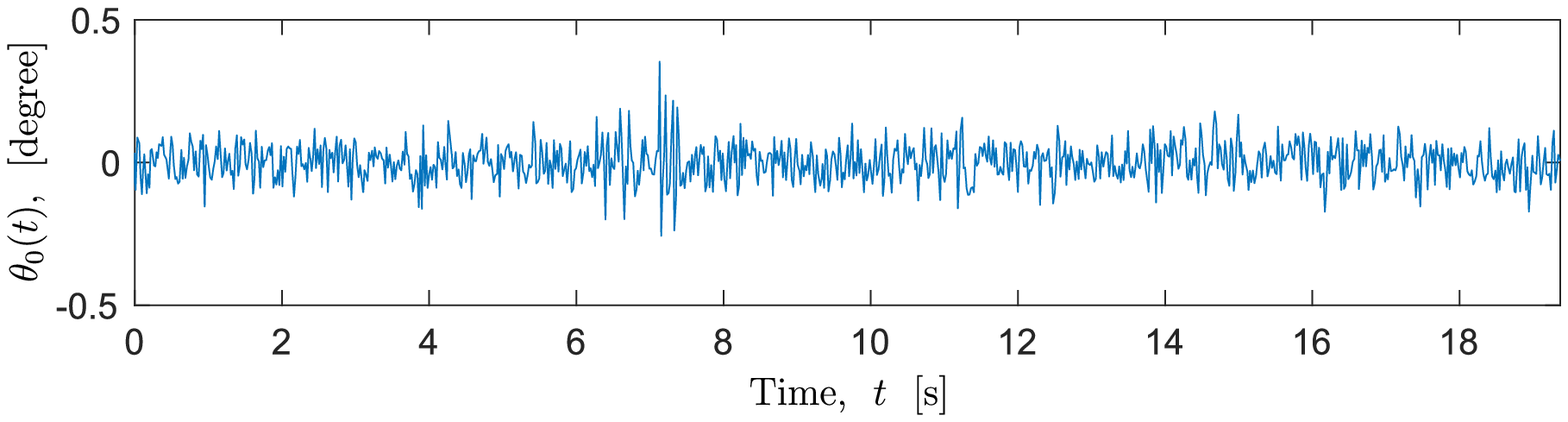}
                \caption{}
            \end{subfigure}~
            \begin{subfigure}[b]{.5\columnwidth}
                \centering
                \includegraphics[width=1\columnwidth,draft=false]{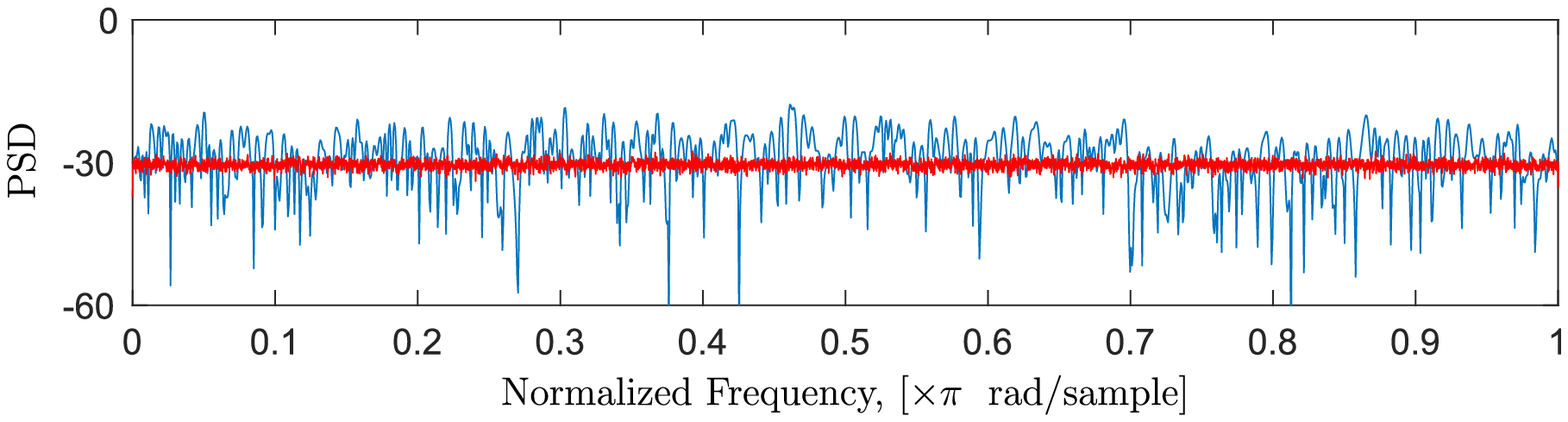}
                \caption{}
            \end{subfigure}\\
            \begin{subfigure}[b]{.5\columnwidth}
                \centering
                \includegraphics[width=1\columnwidth,draft=false]{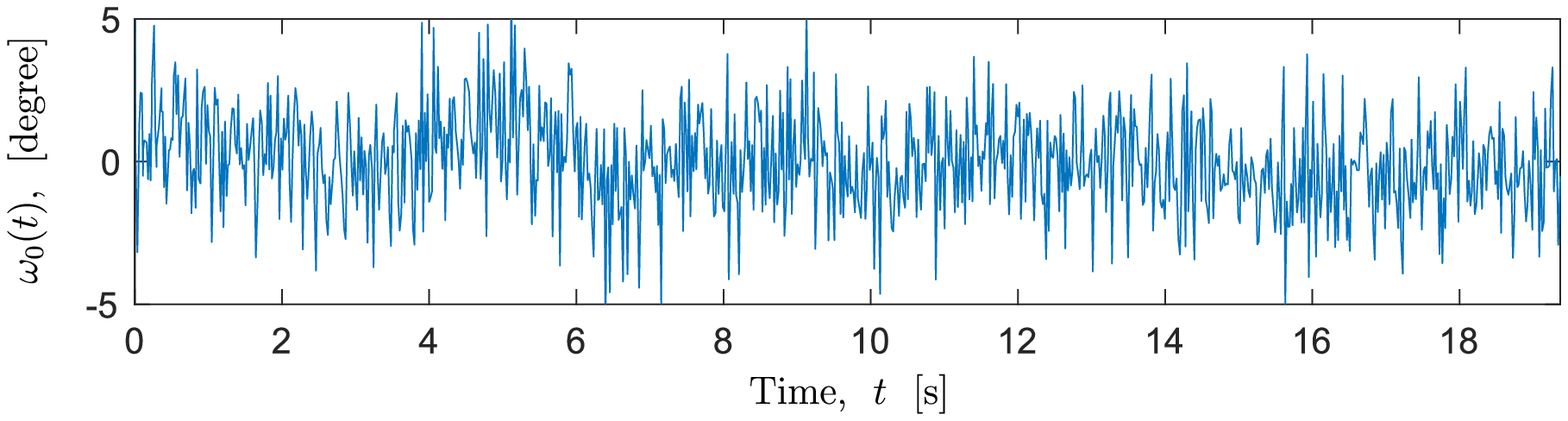}
                \caption{}
            \end{subfigure}~
            \begin{subfigure}[b]{.5\columnwidth}
                \centering
                \includegraphics[width=1\columnwidth,draft=false]{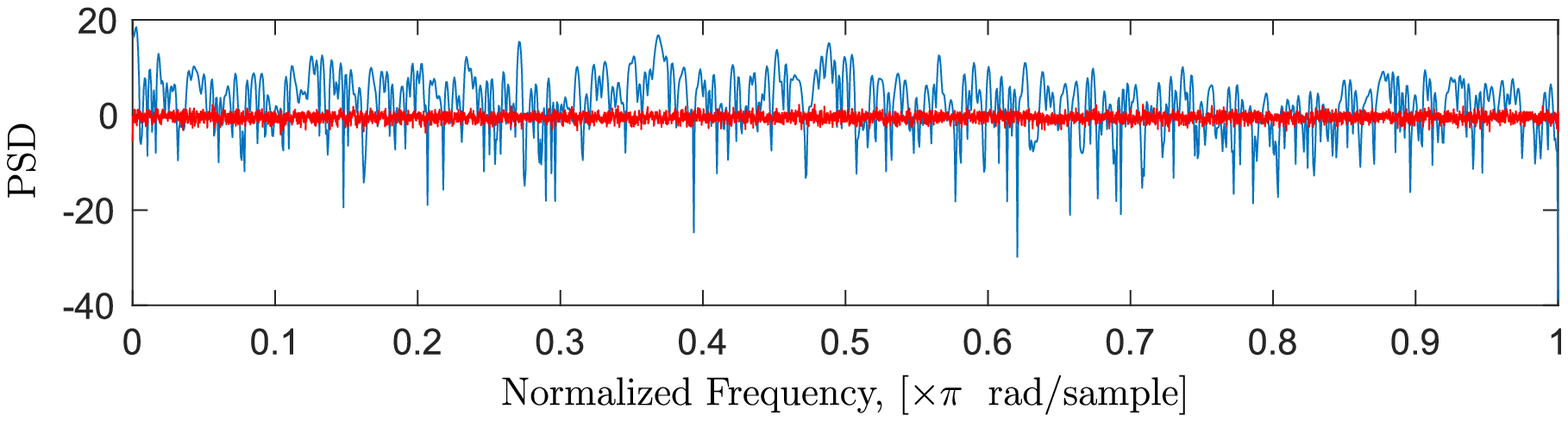}
                \caption{}
            \end{subfigure}\\
            \begin{subfigure}[b]{.5\columnwidth}
                \centering
                \includegraphics[width=1\columnwidth,draft=false]{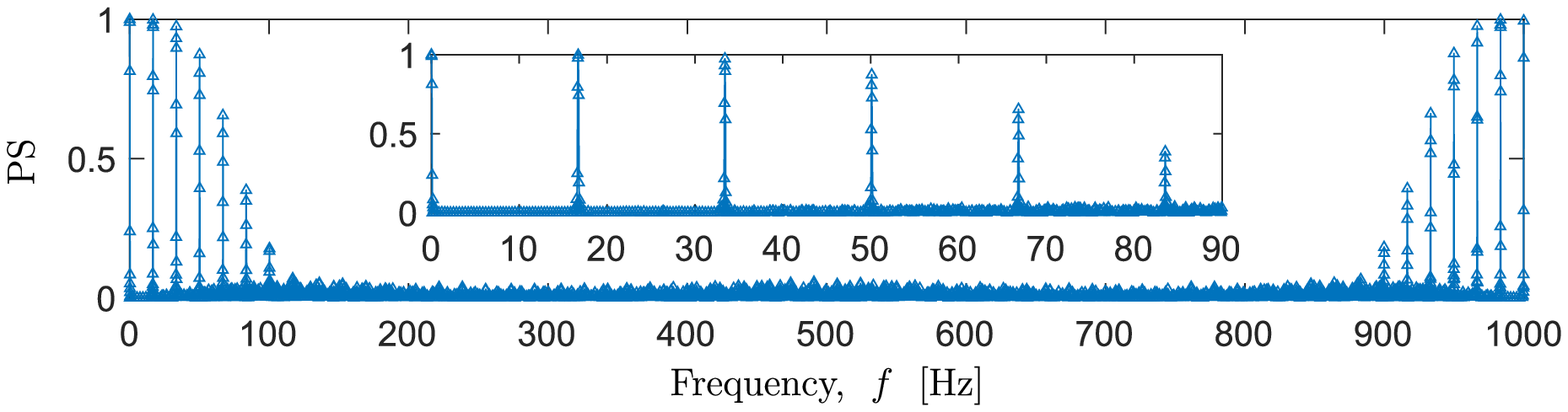}
                \caption{}
            \end{subfigure}
            \caption{Results of measurements with no activity: (a) samples of $\theta_0(t)$, (b) PSD of $\theta_0(t)$ in dB-degree/rad/sample, (c) samples of $\omega_0(t)$, (b) PSD of $\omega_0(t)$ in dB-degree/rad/sample and (e) PS of $w(t)$. \colbluemss{The red curves in (b) and (d) are the average of many generated white noises with the same variances as the respected figures to show the consistency of the model.}}
            \label{fignoises}
        \end{figure}

    Some samples of the results from noise measurements are shown in \figref{fignoises}. Note that the other samples also show similar results as in \figref{fignoises} except \figref{fignoises}(e) as the sampling times are random; thus, the spectrum of the window is also random. The power spectrum (PS) of $w(t)$ is calculated by the LSSA. Our equivalent definitions of the PS and the PSD in uniform time sampling are provided in the Appendix.

    By comparing \figsref{fignoises}(a) and (c), notice that the fluctuations of $\omega_0(t)$ are \colbluemss{stronger}. This fact is supported by their PSDs shown in \figsref{fignoises}(b) and (d), respectively; the PSDs resemble those of white noises, and their variances are $-29.71$ dB\colbluemss{-degree} and $0.11$ dB\colbluemss{-degree}, respectively. Note that these values are determined from their linear scales instead of decibel scales to get a more accurate result. These variances will be used as parameters in our filter for estimating a desired signal from a noisy measurement. Note that the narrow red curves in \figsref{fignoises}(b) and (d) are the average of the PSDs of many generated white noise with corresponding variances. These empirical observations are done to show their resemblances to white noise.

    To gain insight into the window function $w(t)$, recall that for an evenly sampled case, we have Dirac combs as the window functions both in the time and frequency domains. Sampling in the time domain is equivalent to taking a convolution of the signal of interest in the frequency domain with a Dirac comb; therefore, the aliasing after the Nyquist frequency can be seen. For a random sampling, the window function will also be random, and the definition of the Nyquist frequency might not exist depending on the characteristics of the random sampling \cite{vanderplas}. Based on \cite{eyer1999}, the Nyquist frequency for random sampling is $\frac{1}{2\kappa}$, where $\kappa$ is the largest factor such that each spacing of the time sampling is an integer multiple of $\kappa$. In our case, $\kappa$ exists, and $\kappa = 1$ ms; hence, the aliasing after $500$ Hz can be seen in \figref{fignoises}(e). 

    Periodic, decreasing spikes around every $16.7$ Hz can be seen in \figref{fignoises}(e). If we have a sinusoid with a spike in the frequency domain, convolving with such a window function can be mistaken as harmonics. Such a phenomenon is termed as a \textit{partial-aliasing} in \cite{vanderplas}, and the window function from our data measurements falls into the periodic nonuniform sampling class. This class of window function commonly occurs in the instrumentations as the trigger signal periodically changes (see \cite{babu2010} and \cite{baisch1999}), and this phenomenon is relevant in our case. 

    Up to this point, it can be concluded that the noise in our measurement resembles white noise whose variances are $-29.71$ dB-degree for $\theta_0(t)$ and $0.11$ dB-degree for $\omega_0(t)$, respectively. We can also expect that there will be partial-aliasing in the frequency domain and an aliasing after $500$ Hz in the PSD of our data measurement. 

    \subsection{Data Measurements}
    Moving forward from the discussion of noise measurements, we are now ready to discuss our data measurements. The \colredmss{fluctuations} in \figref{figsamples} is also shown in many tremor-related publications, such as \cite{gresty1990} or \cite{dick2017}. Note that the fluctuation not only occurs for subjects who suffer from severe tremors or Parkinson's disease, but the same thing also occurs for healthy subjects with a relatively smaller amplitude as the tremor is defined as rapid involuntary oscillations of parts of the human body \cite{dick2017}. One of the frequently used models to describe such a phenomenon is the harmonic RP, i.e., sinusoids in white noise and with random phases \cite{gresty1990}. Therefore, our work will be based on the harmonic RP. An additional advantage is that it is relatively easy to analyze in terms of \colbluehh{peak} detection, see the appendix for a more detailed explanation. 

    \figref{figressamples} shows the power \colbluehh{spectra} that are estimated by LSSA and the interpolation methods. The dashed line in \figref{figressamples} shows a false alarm probability of $0.1\%$ denoting that the peaks that are above the line have $0.1\%$ probability of an error being mistaken as the real peak. Both PSs are calculated with the same Nyquist frequency, i.e., $500$ Hz. Since the results for the $\omega$, i.e., $m_{\omega|\tilde{\omega}}(t)$, show the same characteristics, the results are not presented here. 

      \begin{figure}
            \centering
            \begin{subfigure}[b]{.5\columnwidth}
                \centering
                \includegraphics[width=1\columnwidth,draft=false]{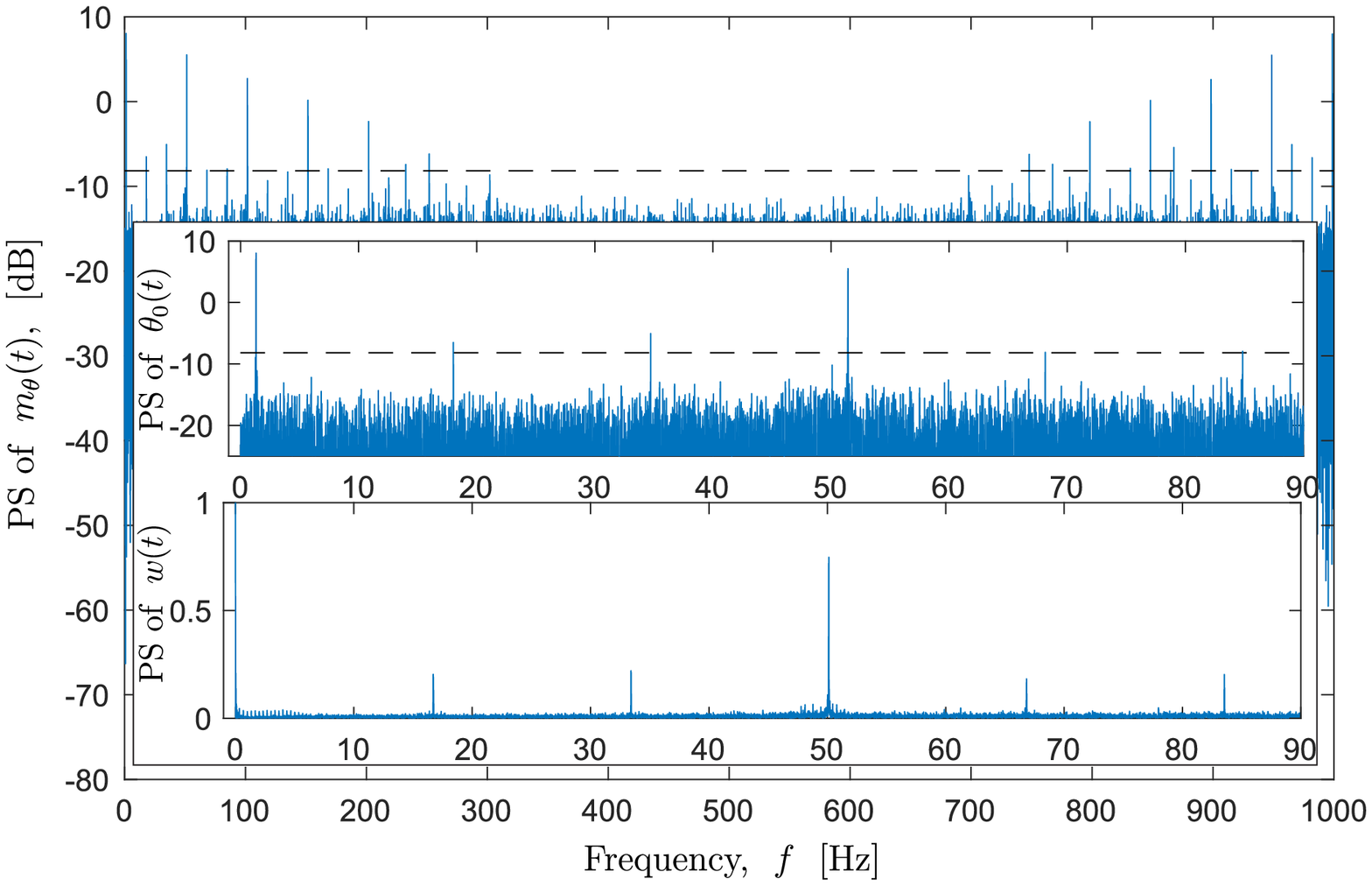}
                \caption{}
            \end{subfigure}~
            \begin{subfigure}[b]{.5\columnwidth}
                \centering
                \includegraphics[width=1\columnwidth,draft=false]{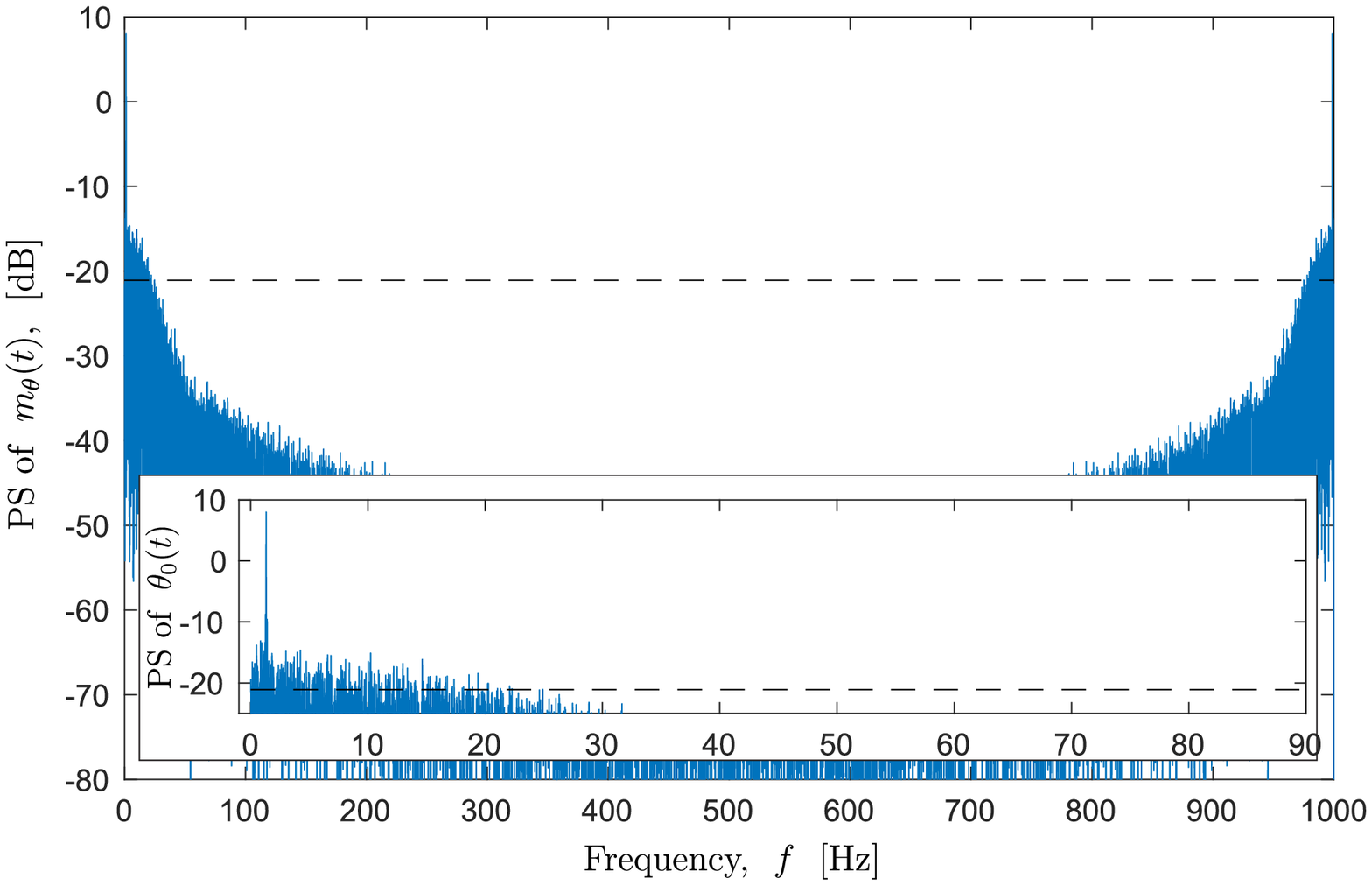}
                \caption{}
            \end{subfigure}
            \caption{Power spectrums of the noisy measurements $m_\theta(t)$ from one of the participants, where: (a) the LSSA is applied, and (b) the interpolation is applied.}
            \label{figressamples}
        \end{figure}

    As previously mentioned, both aliasing and partial-aliasing can be seen in \figref{figressamples}(a). The lower inset \figref{figressamples}(a) shows the PS of the window function. Notice that the frequency gaps of the reliable peaks, i.e., the peaks that are located above the false alarm probability line, have the same frequency gaps as those of the PS of the window function. Therefore, a cleaning algorithm must be performed to detect the real peaks. 

    Our empirical observation on the PSDs shows that the noise floor is also increasing. Therefore, we assume that there are \colbluehh{two white noise sources}, i.e., one of them comes from the noise of the measurement, and the other one is inherent from the participant's unsteady hand. The noise floor is also detected in \cite{dick2017}, and it also agrees with the model in \cite{gresty1990}. In this paper, we will only filter the measurement noise that is explained in the previous subsection. 
    


    The effect of the linear interpolation can be seen in \figref{figressamples}(b), which is a typical smoothing effect. Interpolating unevenly sampled data is equivalent to filtering out a high frequency component. Recall that the cutoff frequency of the linear interpolation is inversely proportional to the number of upsampling ratios. Since our random sampling can range from an order of milliseconds to hundreds of milliseconds, the upsampling ratio is very high; thus, the cutoff frequency of the linear interpolation is \colbluemss{quite low}. \colredmsss{Therefore, quantities, such as ACF,} that are derived from the filtered data will be significantly affected. For more detailed expositions of why the interpolation method is unfavorable than the LSSA, the readers could consult to \cite{vaniceklssa,vanderplas,scargle1982,baisch1999} or \cite{babu2010}.

    Before discussing our model, an independency test between $m_\theta(t)$ and $m_{\omega|\tilde{\omega}}(t)$ in our measurement data needs to be \colredms{carried out}. Note that the random sampling is not an issue here because the test is performed for both $m_\theta(t)$ and $m_{\omega|\tilde{\omega}}(t)$ which have the same sampling time. A test given by \cite{hsictest} is applied. The output of the test is the p-value with the null hypothesis whether two tested RPs are independent. By calculating the p-value for each measurement, we have a minimum p-value of $0.13$ for the sitting activity and $0.19$ for the walking activity. Recall that the smaller the p-value, the stronger the evidence that the null hypothesis should be rejected. In addition, a p-value equal to $0.05$ is used as a rule-of-thumb; hence, p-value larger than $0.13$ is deemed significant. In other words, both $m_\theta(t)$ and $m_{\omega|\tilde{\omega}}(t)$ can be treated as independent RPs. 


    \colredmss{Based on \cite{gresty1990}, $\theta_0(t)$ is modeled as a harmonic RP in white noise, that is:}
        \begin{align}\label{eqmodel}
            \theta_0(t) = A_\theta \ssin{2\pi f_\theta t + \phi} + v_\theta(t),
        \end{align}
    \noindent where $A_\theta$ is the amplitude, $f_\theta$ is the fundamental frequency of $\theta_0(t)$, $\phi$ is a RV that is uniformly distributed from $-\pi$ to $\pi$ and $v_\theta(t)$ is a white noise with the variance $\sigma_{v_\theta}^2$. \colredmss{We will focus on $\theta_0(t)$, but keep in mind that the same model can also be applied for $\omega_0(t)$, and we provide the final result for $\omega_0(t)$ together with that for $\theta_0(t)$.} 
    
    Recall that the unnormalized ACF of \eqref{eqmodel} is:
        \begin{align}\label{eqacf}
            R_\theta(\tau) = \EX{\theta_0(t) \theta^*_0(t+\tau)} = \frac{A_\theta^2}{2} \ccos{2\pi f_\theta \tau} + \sigma^2_{v_\theta} \delta(\tau),
        \end{align}
    \noindent where $\delta(\cdot)$ is an impulse function. The normalization is taken such that the ACF at \colbluemss{$\tau=0$} is one, i.e., $R_\theta(\tau)/R_\theta(0)$. This model agrees with our observation and will be shown later. Our empirical observation for all of our data measurements shows that one sinusoid is \colbluehh{sufficient}.
    

    It is clear that to estimate $\theta_0(t)$, the noise, $n_{\theta_0}(t)$, and the partial aliasing due to the random sampling need to be filtered. For the former case, we use a Wiener filter \cite{hayesbook}, and for the latter case, we use the CLEAN algorithm, see \cite{robertsclean} and \cite{baisch1999}. The CLEAN algorithm is an iterative method that uses the knowledge of a window function to perform deconvolution in the frequency domain. The output of the algorithm is the spectrum as if it were taken from evenly sampled data. 

    In this paper, a first-order Wiener filter is applied, which is denoted as a finite impulse response (FIR) $F(z) = f_0 + f_1 z^{-1}$. A more complicated filter is also possible, but we want to show that even with a simple optimal filter our model can closely match the experimental data. The polynomial of the filter $F(z)$ can be easily calculated by solving following the Wiener-Hopf equation \colbluemss{\cite{hayesbook}}:
        \begin{align}\label{eqwienerhopf}
            \begin{bmatrix}
                \frac{A_\theta^2}{2} + \sigma_{v_\theta}^2 + \sigma_{n_\theta}^2 & \frac{A_\theta^2}{2} \ccos{2\pi f_\theta \tau} \\
                \frac{A_\theta^2}{2} \ccos{2\pi f_\theta \tau} & \frac{A_\theta^2}{2} + \sigma_{v_\theta}^2 + \sigma_{n_\theta}^2
            \end{bmatrix} 
            \begin{bmatrix}
            f_0 \\
            f_1
            \end{bmatrix}
            =
            \begin{bmatrix}
            \frac{A_\theta^2}{2} + \sigma_{v_\theta}^2  \\
            \frac{A_\theta^2}{2} \ccos{2\pi f_\theta \tau}
            \end{bmatrix}.
        \end{align}
    \noindent All parameters in \eqref{eqwienerhopf} can be obtained by calculating both the PSD and PS of the zero-mean noisy measurement. The values of the parameters will be explained in the next paragraph.

\begin{figure}
            \centering
            \begin{subfigure}[b]{.5\columnwidth}
                \centering
                \includegraphics[width=1\columnwidth,draft=false]{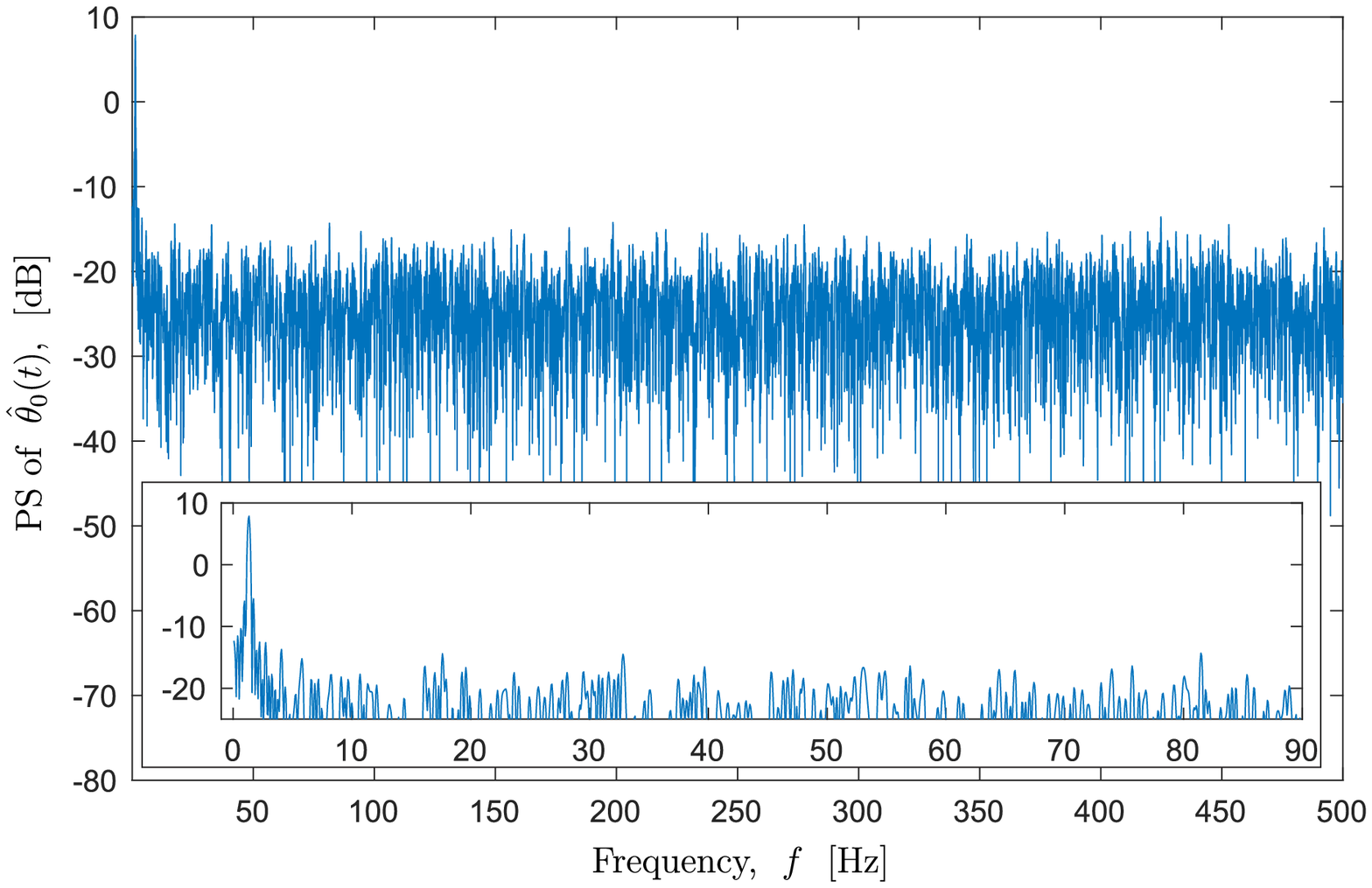}
                \caption{}
            \end{subfigure}~
            \begin{subfigure}[b]{.5\columnwidth}
                \centering
                \includegraphics[width=1\columnwidth,draft=false]{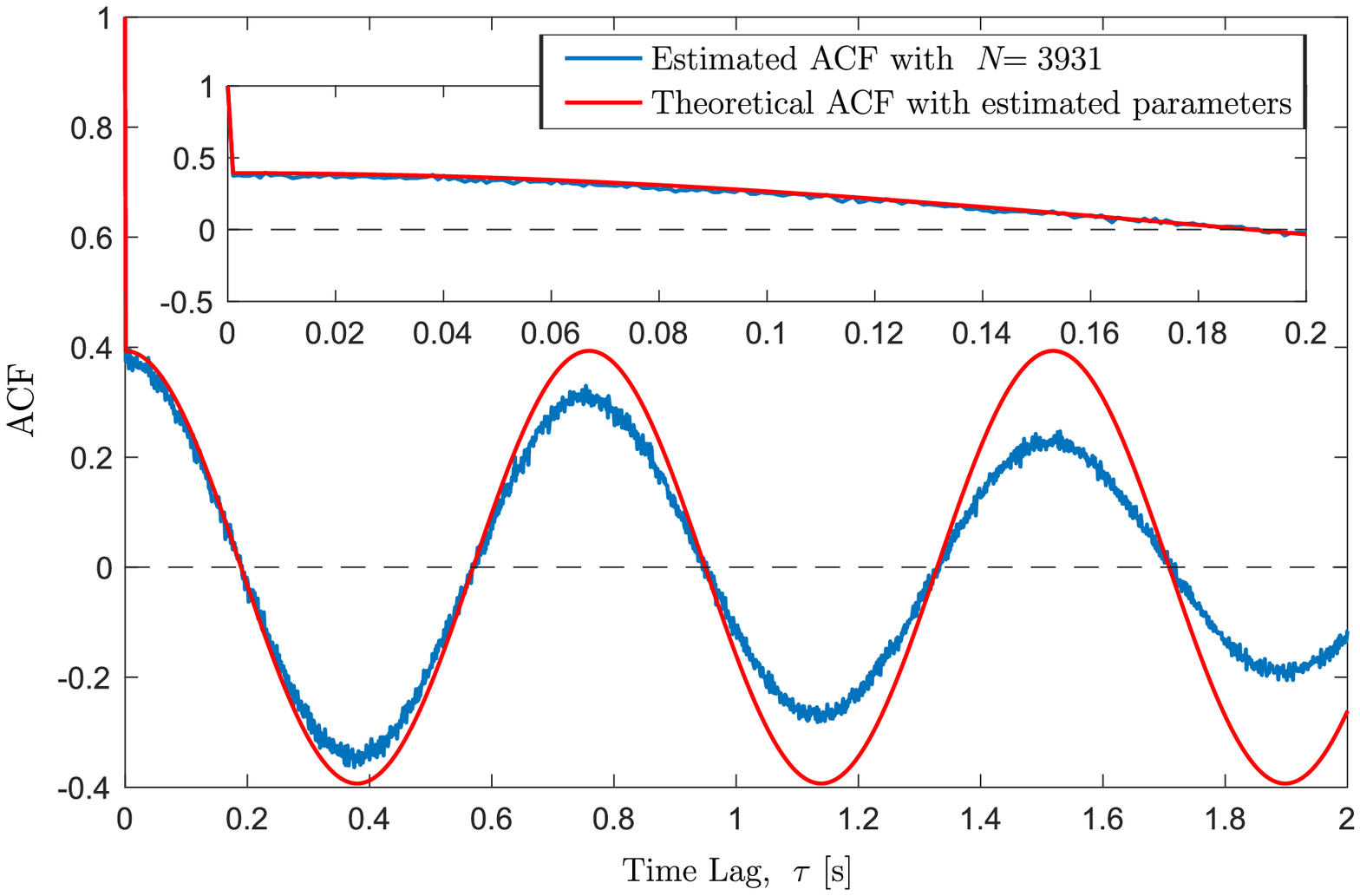}
                \caption{}
            \end{subfigure}
            \caption{(a) A power spectrum of $\hat{\theta}_0(t)$ from \figref{figressamples} after being filtered and (b) an ACF of $\hat{\theta}_0(t)$ from \figref{figressamples}.}
            \label{figpsdestimated}
        \end{figure}

    \figref{figpsdestimated}(a) shows the PS of the estimated $\theta_0(t)$, $\hat{\theta}_0(t)$, after being filtered by both the FIR filter and the CLEAN algorithm. It is clear that the partial aliasing no longer exists. From \figref{figpsdestimated}(a), the peak can be detected at $f_{\theta} = 1.32$ Hz, and the amplitude is $A_\theta = 3.56^\circ$. 




    Based on \eqref{eqacf}, $\sigma_{v_\theta}$ can be estimated as:
        \begin{align}\label{}
            \sigma^2_{v_\theta} \approx \frac{R_\theta(0) A_\theta^2 \left(1- R_\theta(\epsilon)/R_\theta(0)\right)}{2 R_\theta(\epsilon)},
        \end{align}
    \noindent for a small $\epsilon$. \figref{figpsdestimated}(b) shows the normalized ACF that is calculated by taking the inverse Fourier transform of the PSD of $\hat{\theta}_0(t)$. For \figref{figpsdestimated}(b), $R_\theta(\epsilon)/R_\theta(0)$ is $0.4$ by inspection; therefore, $\sigma^2_{v_\theta}$ is approximately $9.05^\circ$ ($\sigma_{v_\theta}^2 = 9.78$ dB-degree). 


    By plugging in all these estimated parameters, the theoretical ACF can be calculated and it is shown in \figref{figpsdestimated}(b). Note that the estimated ACF is biased, and is calculated by taking the inverse Fourier transform from the experimental data; it means that the ACF is decreasing over the time lag. Recall that in estimating the ACF, the biased estimation is preferable over the unbiased one, as the unbiased one might have the normalized ACF value greater than one, which is not true (see \cite{craymerdissertation} for an example). \colredmss{From \figref{figpsdestimated}(b), notice that the frequency $f_\theta$ from \textit{one of the measurement samples} and our model shows a similarity, especially, in the low region of the time lag as it is the region of interest from the perspective of the wireless communication community. \colredmsss{Such} similarity is also \colredmsss{seen} for other measurement samples.}

    To conclude, we first perform the CLEAN algorithm and the Wiener filter to estimate the values of $A_\theta$ and $f_\theta$. By taking the inverse Fourier transform of the PSD that is the output of the CLEAN algorithm, we estimate the value of $\sigma_{v_\theta}$. These procedures are performed for all of our data measurements. By taking the average from all estimated values of the parameters, we obtain the values as shown in Table~\ref{tabparams}.

        \begin{table}
            \centering
            \caption{Average values of the estimated parameters.}
            \label{tabparams}
            \scalebox{.9}{
			\begin{tabular}{l|l|l|l|l}
            \multirow{2}{*}{Parameters} & \multicolumn{2}{c|}{Sitting Activity} & \multicolumn{2}{c}{Walking Activity} \\ \cline{2-5} 
                                        & $\theta_0(t)$      & $\omega_0(t)$      & $\theta_0(t)$      & $\omega_0(t)$      \\ \cline{1-5}
            $A$ ($^\circ$)                         & $1.88$                   & $1.31$                    &    $3.22$                &     $3.15$               \\ \cline{1-5}
            $f$ (Hz)                         & $0.67$                  & $1.46$                    &       $1.86$             &     $1.71$               \\ \cline{1-5}
            $\sigma_v$ ($^\circ$)                  & $5.91$                   & $3.22$                    &     $7.59$               &      $9.48$              \\ \cline{1-5}
            \end{tabular}
            }
        \end{table}

    Based on the value of $\sigma_v$ from Table~\ref{tabparams}, it is confirmed that the fluctuation of the experimental results for the walking activity is higher. The ACF will reach $0$ for the first time at $t = 1/(4 f)$, and the lowest value is $0.13$ s. Since the delay spread of the CIR $h(t)$ is typically in the order of nanoseconds \cite{infraredkahnbarry}. Therefore, $h(t)$ can still be considered slowly varying since the change of orientation is highly correlated in the timescale of nanoseconds. 
    


\colredmsss{Let's now define $\omega$ and $\theta$ as RVs whose realizations are chosen randomly and by taking evenly-spaced samples from their RP models without the measurement noise, i.e.:
    \begin{align*}
        \omega = \tilde{\omega} + \omega_0(n T_\text{s}),\\
        \theta = \mu_\theta + \theta_0(n T_\text{s}),
    \end{align*}
    \noindent where $n \in \mathbb{N}$ and $T_\text{s}$ is a sampling time. For $\omega$, Table~\ref{tabparams} and the results in \cite{soltanitcom} and \cite{zhihongorientation} suggest that $\omega$ can be accurately modeled as a uniformly distributed RV as $\omega_0$ has a relatively small variance compared to $\tilde{\omega}$, which is uniformly distributed in $(-\pi, \pi]$. For modeling conditional azimuth angle (i.e., $\omega|\tilde{\omega}$), $\omega_0(t)$ can be generated based on the harmonic RP discussed above using, for example, a Gaussian white noise.}

\begin{figure}
            \begin{center}
                \includegraphics[width=.5\columnwidth,draft=false]{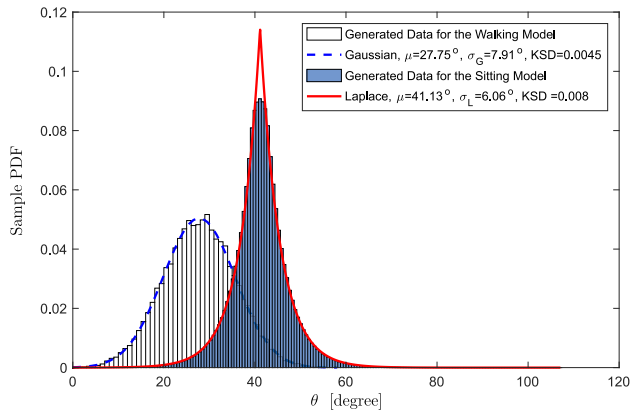}
                \caption{\colredms{Samples PDFs of $\theta$ with evenly spaced sampling and the fitted distributions.}}
                \label{figpdfuniformsampling}
            \end{center}
        \end{figure}

    Regarding the RV $\theta$, it is shown in \cite{ardimaswcnc} \colredms{and \cite{soltanitcom}} that the PDF of $\theta$ for the sitting activity is closer to the Laplace distribution, and it is closer to the Gaussian distribution for the walking activity \colredmss{in terms of Kolmogorov-Smirnov distance (KSD).} \colredms{Recall that $v_\theta(t)$ in \eqref{eqmodel} is kept general in the sense that the white noise RP can be generated from any independent and identically distributed RV. We observe that to match it with the RV models in \cite{ardimaswcnc,soltanitcom}, $v_\theta(t)$ \colbluemss{can be} modeled to be $v_\theta(t) = X$, where $X$ is a zero-mean RV that follows either a Laplace distribution or a Gaussian distribution depending on the activity \colbluemss{as shown in \figref{figpdfuniformsampling}}.} 
    
    
    \colredms{The closed-form PDF of $\theta$ is hard to obtain since the derivation involves an integration of a function that has non-elementary functions. That is, the PDF of $\theta$ is the convolution of a Laplace or a Gaussian distribution and an arcsine distribution, which has the term $\propto 1/\sqrt{C-x^2}$, where $C > 0$. Therefore, the moments matching method is used, and the result is shown in \figref{figpdfuniformsampling}. The KSDs are used to measure the differences between \colbluemss{the evenly-spaced generated samples based on our model and the fitting Gaussian or Laplace distributions.} Based on the KSDs, \colbluemss{we can observe that generated samples of $\theta$} as a RV follow either a Laplace or a Gaussian distribution. It is straightforward to calculate the moments of the fitted distribution, e.g., the mean of the Gaussian distribution is $\mu_\text{G} = \mu_\theta$, and the variance is $\sigma^2_\text{G} = \sigma^2_v+ A^2/2$. The same equation is also applied for the Laplace distribution by replacing the subscript `G' with `L' to denote which distribution is being used. \colredmss{It is worth noting that this RP model can be used in conjunction with the generalized random way point model proposed in \cite{soltanitcom} for modeling the mobility of the users in LiFi networks.}}

\section{Fixed Location and Orientation}
    
    In this paper, before discussing the general case where the locations and the directions of the users are random, several specific cases are discussed so that a deeper understanding both in specific and general cases can be obtained in the end. Particularly, we will first discuss the behavior of the CIR and its effect on our OFDM system in terms of the BER and the received SNR to achieve a certain forward error correction (FEC) threshold.  

    \subsection{Channel Impulse Response}

\begin{figure}
                \centering
                \begin{subfigure}[b]{.5\columnwidth}
                    \centering
                    \includegraphics[width=1\columnwidth,draft=false]{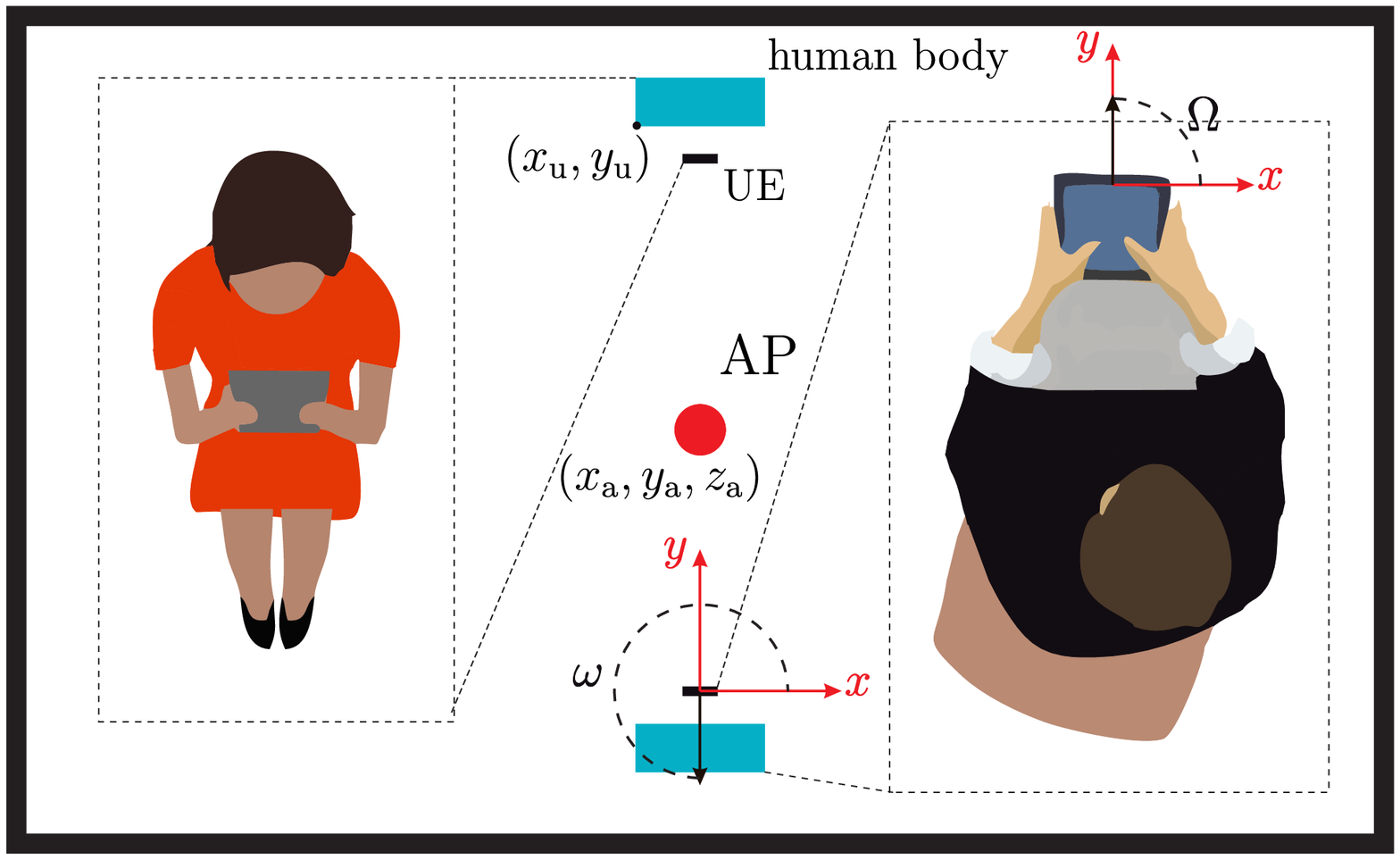}
                    \caption{}
                \end{subfigure}~
                \begin{subfigure}[b]{.5\columnwidth}
                    \centering
                    \includegraphics[width=1\columnwidth,draft=false]{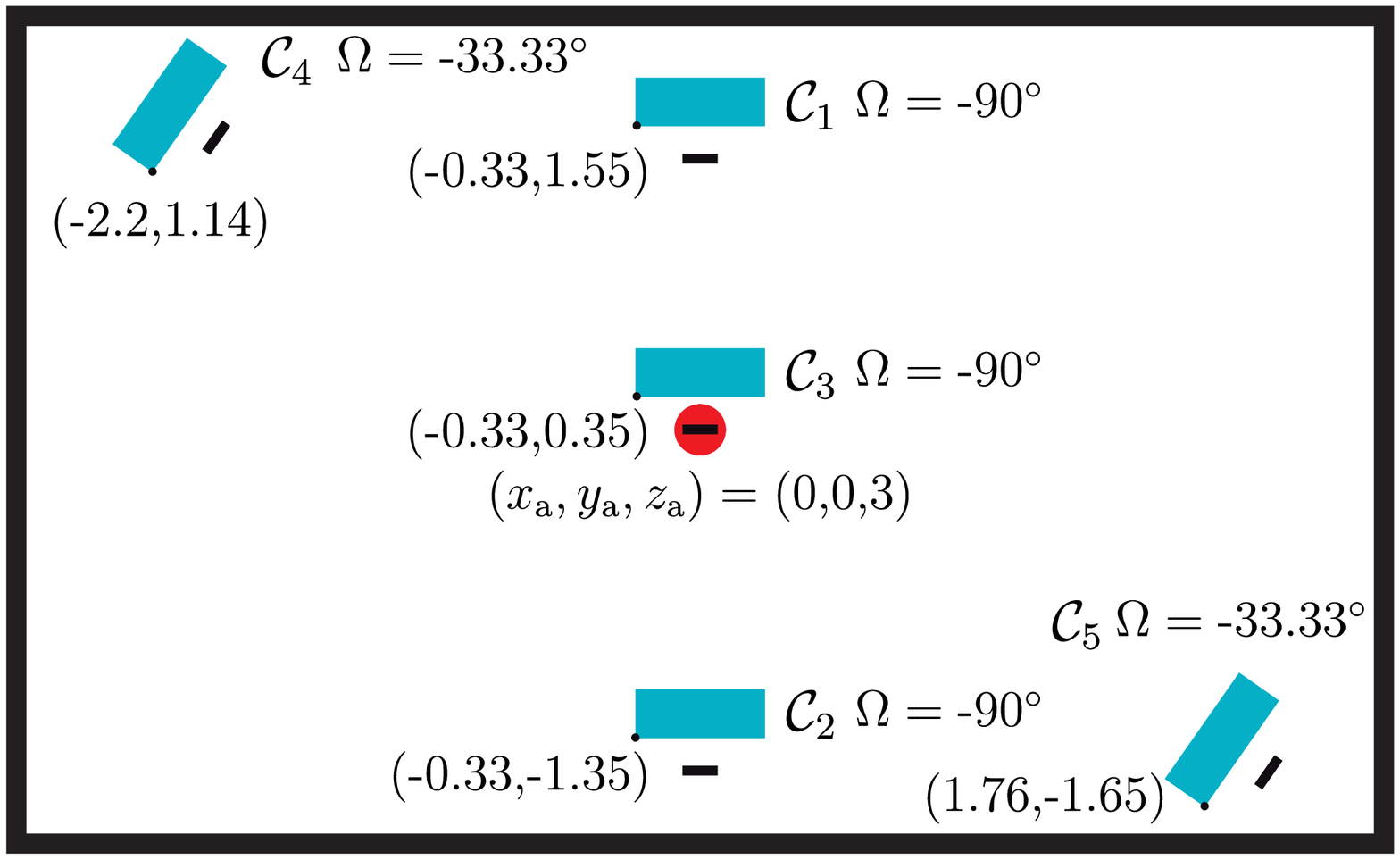}
                    \caption{}
                \end{subfigure}
                \caption{(a) Description of the locations and the directions of users and UEs and (b) configurations of locations and directions of the users and the UEs.}
                \label{angleviewsystemmodel}
            \end{figure}

        In this section, the CIR \colbluemss{behavior} for different configurations will be discussed. \figref{angleviewsystemmodel}(a) shows the parameters to describe configurations of interest in this paper. The location of the AP is denoted by $\left(x_\text{a}, y_\text{a}, z_\text{a} \right)$. Since a human body is modeled as a rectangular prism with dimensions of $L_\text{b} \times W_\text{b} \times H_\text{b}$, its location is represented by one of the locations of its vertices as denoted by $\left(x_\text{u}, y_\text{u} \right)$ without the $z$-axis coordinate. The location of a UE is relative to the location of the user who holds it. In other words, we assume that the users hold the UEs in front of their chests. These are described in the insets of \figref{angleviewsystemmodel}(a). The direction of the UE is denoted by $\omega$, and the direction of the user is assumed to be $\Omega = \omega + \pi$ as depicted in \figref{angleviewsystemmodel}(a).


         As in the experiments, two activities are considered, i.e., the sitting and the walking activities. Note that the term `walking activity' is used throughout the rest of the paper so that it is consistent with the term used in the experiments even though it is more intuitively correct to view it as a case where the user stands while holding the UE with a certain direction. We will use $\theta = 41.13\degr$ for the sitting activity, and $\theta = 27.75\degr$ for the walking activity, which are equal to the means obtained in the experiments. For the walking activity scenario, we define $L_\text{b} = 0.66$ m, $W_\text{b} = 0.2$ m and $H_\text{b} = 1.75$ m \cite{humandimension}. The UE is placed $0.35$ m in front of the user's chest, whose height is calculated as $1.4$ m tall. For the sitting activity scenario, the only differences are that $H_\text{b} = 1.25$ m, and the UE's height is $0.9$ m. 

         The room dimensions are assumed to be $L = 5$ m, $W = 3.5$ m and $H = 3$ m. The reflectivities of the walls are assumed to be $\rho = 0.3$, the reflectivity of the ceiling is $\rho = 0.69$ and the reflectivity of the floor is $\rho = 0.09$ \cite{barrykahnsim}. The reflectivities of the surfaces of a human body are assumed to be $\rho = 0.6$, which is the reflectivity of a cotton fabric \cite{cottonfabric}. Note that a smaller rectangular prism can also be \colbluemss{used} to model the head of \colbluemss{a human to distinguish it from the main body}. However, \colredmss{this is not included in this paper}. In addition, the reflectivity of human skin is also around $0.6$ \cite{humanskin}. Therefore, we only model the hair of a human as the upper surface of the rectangular prism, and its reflectivity is $\rho = 0.9$ \cite{propertieshair}. In addition, $m = 1$, $\FoV = 90\degr$, $A_\text{Rx} = 1$ cm$^2$ are used and the resolution of the partition is $10$ to generate the CIR. 


        For this specific case, $5$ different configurations are chosen as denoted by $\mathcal{C}_i$ for the \ith configuration, see \figref{angleviewsystemmodel}(b). These configurations are picked to describe several scenarios of interest which will be explained later. The location of the AP is $(0,0,3)$, i.e., it is located on the ceiling. Given the dimension of a user and the relative position of a UE to the user, each configuration can be only represented by the location of the user and its direction. For example, for $\mathcal{C}_1$, the user is located at $(-0.33, 1.55)$ and his direction is $\Omega = -90\degr$, i.e., the direction of the UE is $\omega = 90\degr$. 

        The configuration $\mathcal{C}_1$ represents the case where the LOS link still exists with $\psi = 64.62\degr$ for the walking activity and $\psi = 70.87\degr$ for the sitting activity. The configuration $\mathcal{C}_2$ represents the case where the LOS link is blocked by the user for the walking activity, but the LOS link exists for the sitting activity with $\psi = 2.14\degr$. The configuration $\mathcal{C}_3$ represents the case where the UE is located underneath the AP. The configuration $\mathcal{C}_4$ represents the case where the user is located at the corner of the room and the LOS link still exists for both activities with $\psi = 72.89\degr$ for the walking activity and $\psi = 80.58\degr$ for the sitting activity. The configuration $\mathcal{C}_5$ represents the case where the LOS link is blocked. However, unlike in $\mathcal{C}_2$, there are two walls near the UE.


            \begin{figure}
                \centering
                \begin{subfigure}[b]{.5\columnwidth}
                    \centering
                    \includegraphics[width=1\columnwidth,draft=false]{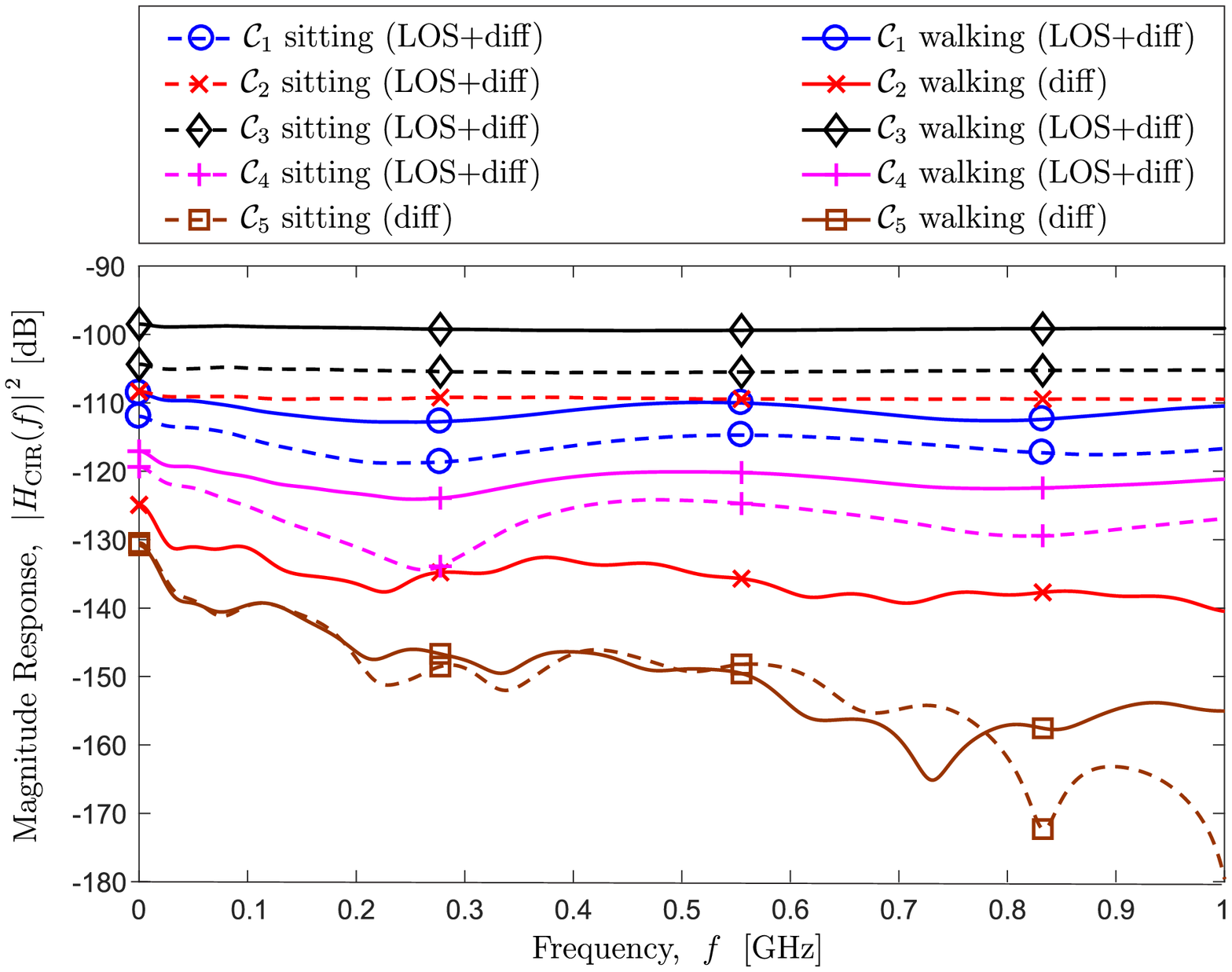}
                    \caption{}
                \end{subfigure}~
                \begin{subfigure}[b]{.5\columnwidth}
                    \centering
                    \includegraphics[width=1\columnwidth,draft=false]{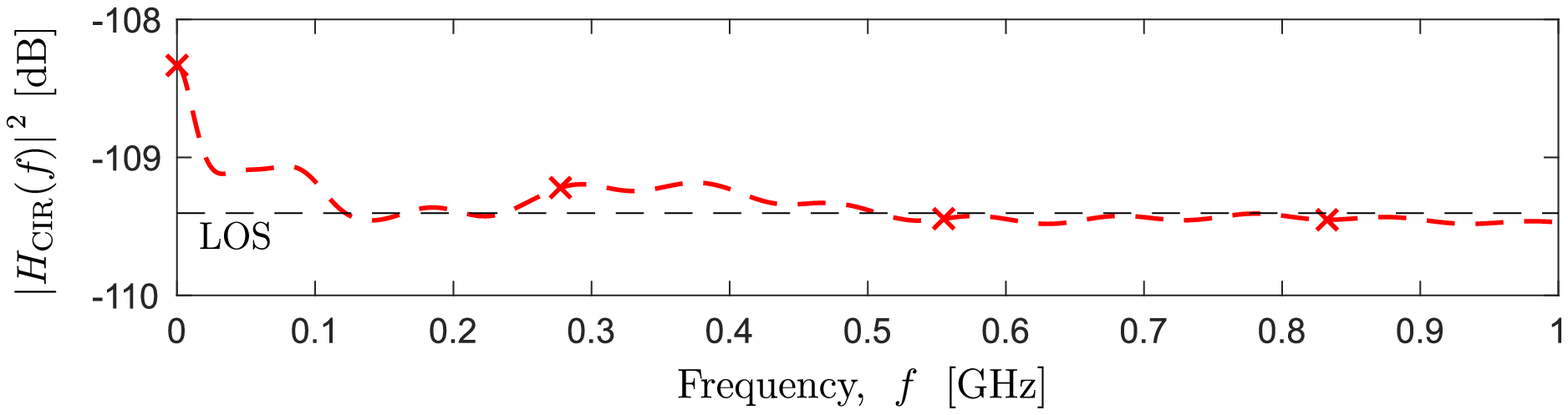}
                    \vspace{1.9cm}
                    \caption{}
                \end{subfigure}
                \caption{(a) Magnitude responses of $\HX{CIR}$ with configurations depicted in \figref{angleviewsystemmodel}(b) and (b) magnitude responses of $\HX{CIR}$ for $\mathcal{C}_2$ and the sitting activity. \colredms{The terms sitting and walking are used here to relate them with our previous discussions, while the physical meanings of them are merely that the person holding the UE sits or stands and, the UE is oriented to the mean value.}}
                \label{figcirfreqresp}
            \end{figure}

        \figref{figcirfreqresp} shows the magnitude response of the CIR with configurations described in \figref{angleviewsystemmodel}(b). It is obvious that the configuration $\mathcal{C}_3$ gives a higher channel gain compared to the others due to its shortest distance. Recall that the participants tend to tilt the UEs to the front during the walking activity than that during the sitting activity. Therefore, the incident angle $\psi$ is smaller; hence, the channel gain is higher for the walking activity. The results for $\mathcal{C}_1$ are comparable with those for $\mathcal{C}_4$. With full $\FoV$, the channel gain for $\mathcal{C}_4$ is $10$ dB lower than that for $\mathcal{C}_1$. It will be shown later how this difference affects the performance of our OFDM system. 
        

        The results for $\mathcal{C}_2$ show the case when the user is located near the wall. The channel gain difference between the results for the sitting activity (the LOS link exists) and the walking activity (the LOS link is blocked) varies between $20$ to $30$ dB. This result is consistent with the result of the shadowing effect in \cite{infraredkahnbarry}. \figref{figcirfreqresp}(b) shows the zoom-in figure of the result for $\mathcal{C}_2$ and the sitting activity. When LOS link exists, observe that the CIR fluctuates around the DC channel gain of its LOS link, especially, in a higher frequency region. It is also interesting to note that the channel gain for the sitting activity with $\mathcal{C}_2$ is slightly higher than that for the walking activity with $\mathcal{C}_1$. This is due to the incident angle $\psi$ is much smaller; hence, the channel gain is higher as it is proportional to $\ccos{\psi}$, see \eqref{eqGRxTx}. The results for $\mathcal{C}_5$ have the worst channel gain of all. These channel gains will be used to determine the number of CP for our OFDM system.

        As in \cite{downlinkcheng}, we are also interested in the ratio between the power from the LOS link denoted by $P_{\text{LOS}}$ and the total power denoted by $P_{\text{tot}}$. Table~\ref{tabpowerratio} shows the ratio in percentage. Note that to save the space, \colbluehh{the subscripts `w' and `s'} denote the results for the walking and sitting activities, respectively. Contrary to the results in \cite{downlinkcheng} which considers only the case where the UE always faces upward, the power contribution of the LOS link is not always dominant which is shown by the results with the configuration $\mathcal{C}_4$. The results of the power ratio will be generalized in the next section for randomly-located users.

        \begin{table}[]
            \centering
            \caption{The contribution of the power from the LOS link in percentage.}
            \label{tabpowerratio}
            \scalebox{.9}{
            \begin{tabular}{l|l|l|l|l|l|l|l}
                                                      & $\mathcal{C}_{1,\text{s}}$  & $\mathcal{C}_{1,\text{w}}$  & $\mathcal{C}_{2,\text{s}}$  & $\mathcal{C}_{3,\text{s}}$  & $\mathcal{C}_{3,\text{w}}$  & $\mathcal{C}_{4,\text{s}}$  & $\mathcal{C}_{4,\text{w}}$\\ \hline
            $\frac{P_{\text{LOS}}}{P_{\text{tot}}} (\%)$   & 61.53                       & 73.28                       & 88.41                       & 89.78                       & 92.51                       & 40.96                       & 60.24 \\ \hline
            \end{tabular}
            }
        \end{table}

    \subsection{OFDM Performance}

        \colbluehh{In this paper, if the values of the parameters are not mentioned, the values are defined as shown in Table~\ref{tabvalparam}}. Given the values of OFDM parameters, the bandwidth of OFDM symbols is around $28$ MHz. These are chosen based on the value of $f_{\text{c}_{\text{LED}}}$ that ranges between $20$ to $40$ MHz, see \cite{ledmodelflat,ledmodelobrien} and \cite{downlinkcheng}. The number of CPs is chosen based on the worst channel in \figref{figcirfreqresp}, which is $\mathcal{C}_5$. The BER target is chosen such that it is lower than the FEC threshold, which is defined as $P_{\text{b,target}} = 3.8\times 10^{-3}$ \cite{8gbpsislim}, and the responsitivity of the PD is $R = 0.6$ A/W \cite{barrosofdm}.

        \begin{table}[]
            \centering
            \caption{Values of parameters.}
            \label{tabvalparam}
            \scalebox{.9}{
			\begin{tabular}{l|l|l}
                Parameters & Values & Units \\ \hline
                Bit rate ($R_\text{b}$) & 100       & Mbps    \\
                Number of subcarriers $\left( N \right)$ &  128     &  -    \\ 
                Number of used subcarriers $\left( N_\text{u} \right)$ &  108     &  -    \\ 
                Number of CPs $\left( N_\text{cp} \right)$ &  7     &  -    \\ 
                Modulation order $\left( M \right)$ &  16     &  -    \\ 
                Cutoff frequency of the LED $\left( f_{\text{c}_{\text{LED}}} \right)$ &  20 or 40     &  MHz    \\ 
                BER target $\left( P_{\text{b,target}} \right)$ &  $3.8\times 10^{-3}$     &  -    \\ 
                Responsitivity of the PD $\left( R \right)$ &  $0.6$     &  A/W    \\ \hline 
            \end{tabular}
            }
        \end{table}

\begin{figure}[!t]
            \begin{center}
                \includegraphics[width=.5\columnwidth,draft=false]{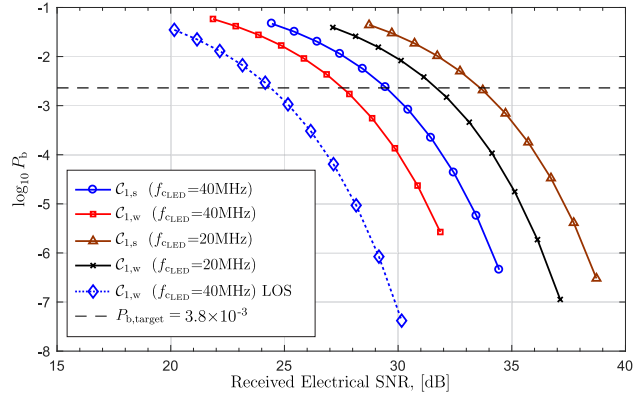}
                \caption{BER comparison for $\mathcal{C}_1$ where the subscript `s' denotes sitting and `w' denotes walking. The solid or dashed lines show the theoretical results, and the markers show the simulation results.}
                \label{figberc1}
            \end{center}
        \end{figure}

        Following \cite{infraredkahnbarry} and \cite{barrosofdm}, a received electrical SNR is defined as:
            \begin{align}\label{eqelecsnr}
                \text{SNR} = \frac{ R^2 \HX{CIR}^2(0) P_\text{t}^2 }{N_0 f_s/2},
            \end{align}
        \noindent where $P_\text{t}$ is the average optical transmitted power. The benefit of this definition is that the performances of different modulation schemes can be fairly compared \cite{barrosofdm}. The SNR that is required to achieve $P_\text{b} = P_\text{b,target}$ is denoted by SNR$_\text{t}$. The BER comparisons for $\mathcal{C}_1$ is depicted in \figref{figberc1}. \colredmss{The result follows our intuition that the multipath propagation makes the BER worse. In addition, as seen in \figref{figberc1}, given the same rate and the same bandwidth of the OFDM symbols, decreasing the bandwidth of the LED such that it is less than bandwidth of the OFDM symbols will also decrease the BER performance. Note that not all BER curves will be presented in the figure since many configurations are investigated in this paper. Instead, only \colredms{the SNR values} at $P_\text{b} = P_\text{b,target}$ are discussed in this paper, and they are given in Table~\ref{tabsnrtarget}.}
        
        We are also interested in investigating the effect of higher order reflections. In other words, the effect of ignoring the diffuse link as in many published works, e.g., \cite{ardimaswcnc,wangtilted,wangber} or \cite{erogluorientation}, will be discussed. In terms of the BER target, we focus on the target SNR penalty, i.e., SNR$_\text{t,penalty}$, which is defined as the difference between SNR$_\text{t}$ with both links and SNR$_\text{t}$ with the LOS link only. The values of SNR$_\text{t,penalty}$ can range from $0.58$ to $10.3$ dB for the channel in \figref{figcirfreqresp}. Notice that the flatter the channel, the smaller the penalty; hence, $\mathcal{C}_{4,\text{s}}$ has the biggest penalty, and the smallest penalty is achieved for $\mathcal{C}_{3,\text{w}}$. The flatness of the CIR is highly correlated to the power ratio as shown in Table~\ref{tabpowerratio}. We also observe that the power ratio of $75\%$ is related to the SNR penalty around $3$ dB. \colbluehh{For system designers who calculate a power budget of a LiFi system, this penalty might significantly affect the whole system cost.}

        \begin{table*}[]
            \centering
            \caption{Comparison of the received electrical SNR (in dB) at $P_{\text{b,target}}$, which is denoted by SNR$_\text{t}$, and the SNR penalty when the diffuse channel is neglected, which is denoted by SNR$_\text{t,penalty}$.}
            \label{tabsnrtarget}
            \resizebox{\linewidth}{!}{
            \begin{tabular}{l|l|l|l|l|l|l|l|l|l|l|l|l|l|l|l|l|l|l|l|l}
            $f_{\text{c}_{\text{LED}}}$   & \multicolumn{10}{c|}{40 MHz}      & \multicolumn{10}{c}{20 MHz} \\ \hline
            $\mathcal{C}$                 & $\mathcal{C}_{1,\text{s}}$  & $\mathcal{C}_{1,\text{w}}$  & $\mathcal{C}_{2,\text{s}}$  & $\mathcal{C}_{2,\text{w}}$  & $\mathcal{C}_{3,\text{s}}$  & $\mathcal{C}_{3,\text{w}}$  & $\mathcal{C}_{4,\text{s}}$  & $\mathcal{C}_{4,\text{w}}$  & $\mathcal{C}_{5,\text{s}}$  & $\mathcal{C}_{5,\text{w}}$  & $\mathcal{C}_{1,\text{s}}$  & $\mathcal{C}_{1,\text{w}}$  & $\mathcal{C}_{2,\text{s}}$  & $\mathcal{C}_{2,\text{w}}$  & $\mathcal{C}_{3,\text{s}}$  & $\mathcal{C}_{3,\text{w}}$  & $\mathcal{C}_{4,\text{s}}$  & $\mathcal{C}_{4,\text{w}}$  & $\mathcal{C}_{5,\text{s}}$  & $\mathcal{C}_{5,\text{w}}$   \\ \hline
            SNR$_\text{t}$         & 29.49                       & 27.56                       & 25.33                       & 33.78                       & 25.31                       & 25.01                       & 34.73                       & 29.54                       & 41.45                       & 38.97                       & 33.61                       & 31.70                       & 29.58                       & 37.43                       & 29.76                       & 29.47                       & 37.63                       & 33.37                       & 46.60                       & 44.41  \\ \hline
            SNR$_\text{t,penalty}$     &  5.06                       &  3.14                       &  0.91                       & -                           &  0.88                       &  0.58                       & 10.31                       &  5.11                       & -                           & -                           &  4.91                       &  3.00                       &  0.90                       & -                           &  1.07                       &  0.78                       &  8.93                       &  4.67                       & -                           & -     \\ \hline
            \end{tabular}}
        \end{table*}

    The effect of the LED's bandwidth can also be seen in the Table~\ref{tabsnrtarget}. For relatively flat channels as in configurations $\mathcal{C}_3$, decreasing the bandwidth increases the SNR penalty. On the other hand, decreasing the bandwidth reduces the SNR penalty for the other configurations. The reason is that the level of fluctuation in the channel plays more significant role than the effect of the subcarriers loss for the configurations other than $\mathcal{C}_3$. More focused studies about these observations are subject of our future works. In the next section, we also discuss the consequence of ignoring the diffuse link in terms of the outage probability, which is \colbluemss{one of the main metric} of the analyses in the network level or the cellular networks.

\section{Random Locations and Orientations}
    
    In this section, random locations and orientations of the UE are assumed. The benefit of this assumption is that the average performance can be obtained as the indoor optical wireless channel highly depends on the geometry of the UE and the AP, see \figref{figcirfreqresp} for an example. Tools from the stochastic geometry are usually used to consider the case with randomly-located users. \colredms{However, even with the LOS link, a closed form expression is hard to obtain and it typically still contains an integral expression, see \cite{downlinkcheng} and \cite{erickcoverage}.} In this section, a semi-analytic approach is used in the sense that the CIRs are generated by Monte Carlo simulations and the SNR target is calculated analytically. 
    
    

    We will assume that the locations of user and UE are uniformly distributed in an indoor room with the same dimensions used in the previous section. Since only a single channel is concerned, each realization has a user and a UE. This is further referred to as the binomial point process. As for the orientation, we assume that $\theta$ follows either \colredmss{a Laplace distribution or a Gaussian distribution depending on the activity} as depicted in \figref{figpdfuniformsampling}, and the direction of the users, $\Omega$, follows a uniform distribution. We generated $1,000$ CIRs for each activity and observed that the probability of the LOS link exists is $88.1\%$ for the walking activity and $94.8\%$ for the sitting activity. In this section, the CIRs having the LOS links are used.

\begin{figure}
            \centering
            \begin{subfigure}[b]{.5\columnwidth}
                \centering
                \includegraphics[width=1\columnwidth,draft=false]{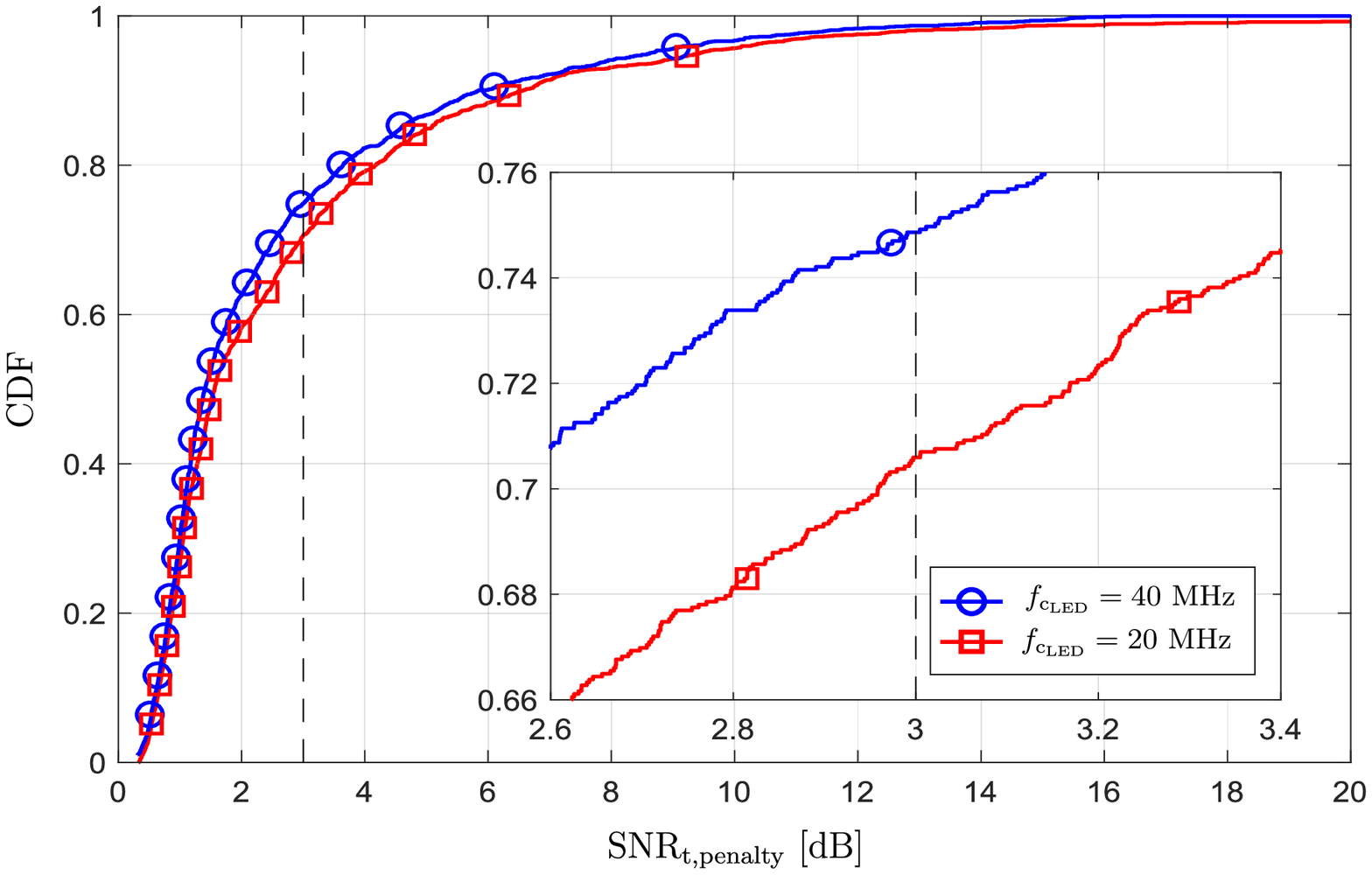}
                \caption{}
            \end{subfigure}~
            \begin{subfigure}[b]{.475\columnwidth}
                \centering
                \includegraphics[width=1\columnwidth,draft=false]{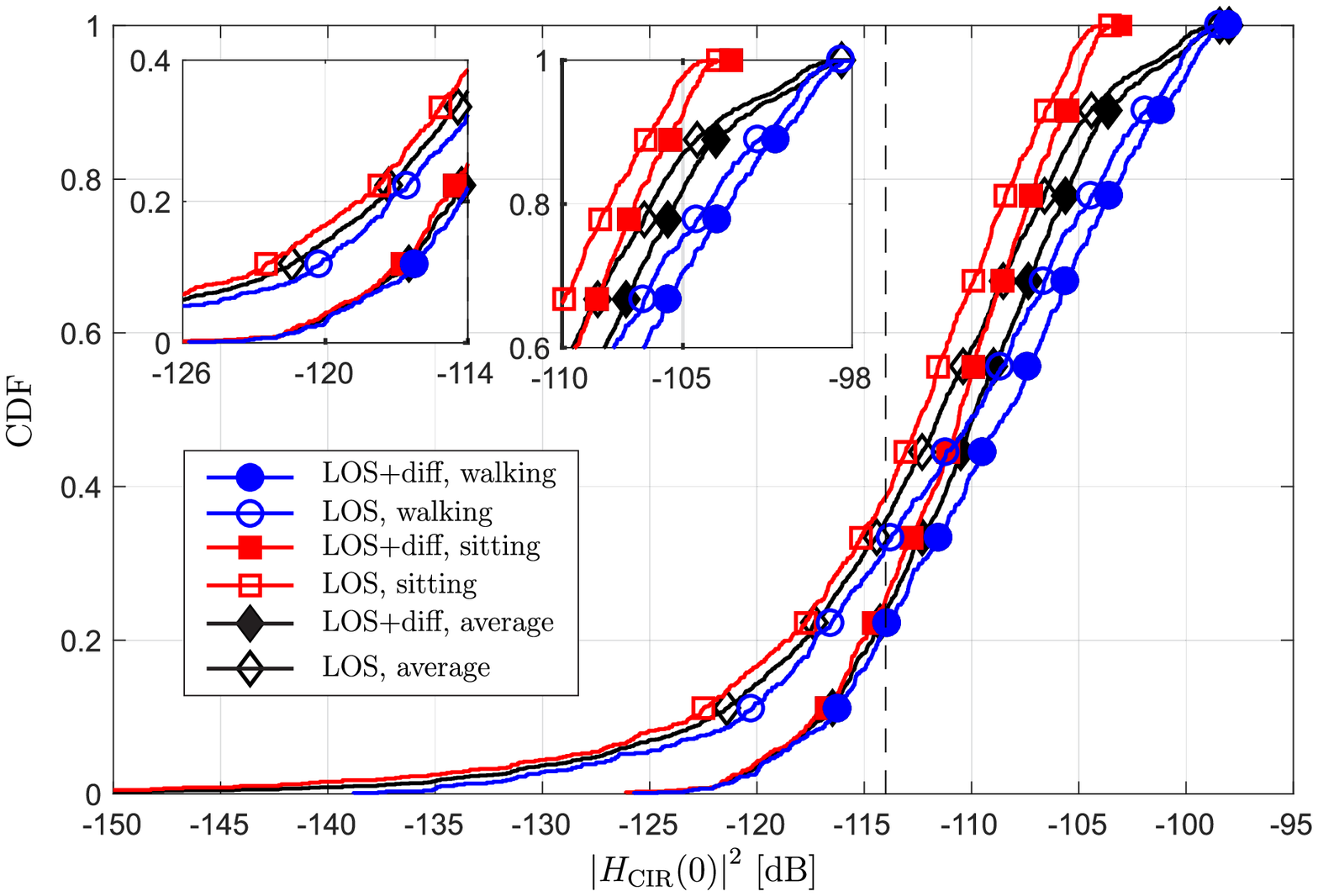}
                \caption{}
            \end{subfigure}
            \caption{(a) CDFs of SNR$_\text{t,penalty}$ with randomly-located, randomly-oriented UEs and (b) CDFs of $|\HX{CIR}(0)|^2$ with randomly-located, randomly-oriented UEs.}
            \label{figsnrpenaltyrandom}
        \end{figure}

    First, we look at the cumulative distribution function (CDF) of SNR$_\text{t,penalty}$ for this random case. The result is depicted in \figref{figsnrpenaltyrandom}(a). Since the results are quite similar, let's focus on $\PX{\text{SNR}_\text{t,penalty} < 3 \ \text{dB} }$ shown in the inset. $\PX{\text{SNR}_\text{t,penalty} < 3 \ \text{dB} }$ for all configurations in \figref{figsnrpenaltyrandom}(a) are around $0.7$. As previously discussed and shown in Table~\ref{tabsnrtarget}, using wider bandwidth results in a better performance in terms of SNR$_\text{t,penalty}$. Generally, considering that there is $0.3$ probability that SNR$_\text{t,penalty}$ can be greater than $3$ dB, we believe that simply ignoring the diffuse link in BER analysis is \colbluehh{too limiting}. Alternatively, one can simply ignore the diffuse link in calculating the BER performance for a randomly-oriented UE for the case where the UE is located near to the AP since the channel is relatively flat, see the results for $\mathcal{C}_3$ in \figref{figcirfreqresp}.



    In a network-level analysis considering the interference and randomly-located UEs as in \cite{downlinkcheng,ardimasglobecom} or \cite{erickcoverage}, the received power is generally assumed to only come from the LOS link. Understanding the previous discussion, we get an insight that this is correct if the UE is located near the AP. Therefore, we are also interested in justifying the importance of \colredmsss{the} diffuse link in the network-level analysis whose the main metric is the signal-to-interference-plus-noise ratio (SINR). Recall that the SINR is a function of the electrical received power which is proportional to the DC channel gain, i.e., $P_\text{r} \propto |\HX{CIR}(0)|^2$, where $P_\text{r}$ is the average received optical power \cite{infraredkahnbarry}. The CDF of $|\HX{CIR}(0)|^2$ is depicted in \figref{figsnrpenaltyrandom}(b).


    Generally, the results for the walking activity are better than that of the sitting activity. The reason is that the distance between the UE and the AP is shorter due to the vertical distance difference between both activities, i.e., the height is $1.4$ m for the walking activity and $0.9$ m for the sitting activity. In addition, we can partition the region of $|\HX{CIR}(0)|^2$ into two parts, i.e., the low region where $|\HX{CIR}(0)|^2 < -114$ dB and the high region where $-114\ \text{dB} \leq |\HX{CIR}(0)|^2$. Note that $-114$ dB is chosen such that the difference of $|\HX{CIR}(0)|^2$ with and without the diffuse link is $3$ dB. Notice that the gap is larger in the low region compared to that in the high region. In \figref{figsnrpenaltyrandom}(b), the maximum gap is around $25$ dB in the low region. In the high region, the gap is narrowing. This is expected since this region is typically obtained when the UE is located near the AP, and \colredmss{the power ratio is relatively high}. For an example, see the CIR for $\mathcal{C}_2$ in \figref{figcirfreqresp}(b). 

    Based on these results, the studies of the interference power can be significantly improved if the diffuse link is incorporated. This is true especially because the interfering transmitters are typically located far from the receiver. Being far from the receiver makes the fluctuation of $\HX{CIR}(f)$ \colbluemss{stronger}, see $\HX{CIR}(f)$ for $\mathcal{C}_4$ and $\mathcal{C}_5$. Using only the LOS link in the studies of the interference can be \colbluehh{considered in} a scenario where the UEs are located near the AP, for example, if the multiple access scheme used is NOMA as in \cite{yapicinoma} and \cite{ericknoma}. \colbluehh{The studies of the interference in NOMA are typically focused on a single cell with multiple users, and the UEs are typically located near an AP within a few meters away, see \cite{ericknoma}}. To relate it with our example, imagine we draw a circle with the center being the AP, and the radius being around $1.5$ m in \figref{angleviewsystemmodel}(b); hence, the frequency responses for $\mathcal{C}_1$ to $\mathcal{C}_3$ are relatively flat, see \figref{figcirfreqresp}.

    \colredms{Another way to interpret results in \figref{figsnrpenaltyrandom}(b) is by viewing it as the outage probability. The relationship between $|\HX{CIR}(0)|^2$ and the outage probability is obvious by looking at \eqref{eqelecsnr}, i.e., the received electrical SNR is proportional to $|\HX{CIR}(0)|^2$. The outage event is typically defined as the event when the received electrical SNR is less than a SNR target to achieve certain BER \cite{aveneauorientation,aveneaubody}. \figref{figsnrpenaltyrandom}(b) shows that a system with both links is significantly better than that with only LOS link, i.e., the outage events less frequently occur if the diffuse link is considered. In other words, reflected signal can help reducing the outage events up to $25$ dB, especially, in the low region of \colbluehh{the DC channel gain}.}  
    


\section{Conclusions}
    This paper focused on modeling the random orientation and applying it in DC-OFDM-based LiFi systems. A series of experiments were conducted with $40$ participants, and $222$ data measurements were obtained while they browsed the Internet and watched streaming videos. The random orientation was modeled as user's slight hand movement and other activities, such as scrolling or typing. In addition, we observed that the sampling time of the sensors were not evenly-spaced. It was also shown that the ACF reached $0$ for the first time when the time lag was $0.13$ s. Compared to the typical delay spread of the optical wireless channel which was in the order of nanoseconds, the CIR could still be assumed to be slowly varying since the random orientation was highly correlated in the order of nanoseconds. 

    In this paper, we also applied the random orientation model to DC-OFDM-based LiFi systems. The RP model with a sinusoid in white noise was used to model the random orientation as an RV. \colredms{The polar angle of the orientation of the UE was modeled as Laplace distribution for the sitting activity and a Gaussian distribution for the walking activity.} In addition, the azimuth angle was modeled as a uniform distribution. Based on the RV model, the error performance of the LiFi system was investigated. We showed that the probability of the SNR penalty which was larger than $3$ dB was $30\%$. In terms of the DC channel gain, it was shown that the penalty could range up to $25$ dB in the low region of the DC channel gain which correlated to the case where the UE was located far from the AP. Therefore, a great care must be taken on when a simplification on reflection could be made.

\appendices
\section*{Appendix: Power Spectrum and Power Spectral Density For Evenly-Spaced Samples}

    Throughout this paper, the spectral analysis is performed based on a LSSA that gives a periodogram, which estimates either power spectral density (PSD) or power spectrum (PS). For simplicity, we can safely discuss and treat the PSD and the PS of the LSSA the same as the evenly-spaced counterpart, see \cite{vaniceklssa,vanderplas,scargle1982,baisch1999} or \cite{babu2010}. However, please bear in mind that the Lomb-Scargle method is used instead, not the conventional Fourier analysis. 
    
    The PSD is mainly used to detect the noise level, and the PS is mainly used to detect the power of a random signal at a certain frequency. Therefore for a discrete random process (RP) $X[n]$ whose length is $N$, its two-sided PS with a rectangular window is expressed by:
        \begin{align*}\label{}
            S_X\left[k\right] = \frac{\left| \sum\limits_{j = 1}^{N} X\left[j\right] \text{e}^{(-2\pi i)(j-1)(k-1)/N} \right|^2 }{N^2},\quad k \in [0,N-1].
        \end{align*}
    \noindent  Meanwhile, its PSD is expressed by:
        \begin{align*}\label{}
            P_X\left[k\right] = \frac{\left| \sum\limits_{j = 1}^{N} X\left[j\right] \text{e}^{(-2\pi i)(j-1)(k-1)/N} \right|^2 }{N}\quad k \in [0,N-1].
        \end{align*}
    \noindent If $X[n]$ is a sinusoid in a white noise with a random phase, i.e., $X[n] = A \ssin{2\pi f_0/F_\text{s} n + \phi} + w[n]$, where $f_0 \neq 0$, $F_\text{s}$ is a frequency sampling, $\phi$ is a random variable (RV) which follows the uniform distribution from $-\pi$ to $\pi$ and $w[n]$ is a zero-mean \colredms{white noise} whose variance is $\sigma^2$, then the estimated power of $X[n]$ at frequency $f_0$ is $2 S_X[N f_0/F_\text{s} ] = \frac{A^2}{2}$. The white noise level is $\sigma^2 = \sum\limits_{j = L}^{U} P_X[j]/(U-L+1)$, where the interval $[L,U]$ is the interval outside the neighborhood of $N f_0/F_\text{s}$.


\bibliographystyle{./IEEEtran}
\bibliography{./report}

\begin{thebibliography}{10}
\providecommand{\url}[1]{#1}
\csname url@samestyle\endcsname
\providecommand{\newblock}{\relax}
\providecommand{\bibinfo}[2]{#2}
\providecommand{\BIBentrySTDinterwordspacing}{\spaceskip=0pt\relax}
\providecommand{\BIBentryALTinterwordstretchfactor}{4}
\providecommand{\BIBentryALTinterwordspacing}{\spaceskip=\fontdimen2\font plus
\BIBentryALTinterwordstretchfactor\fontdimen3\font minus
  \fontdimen4\font\relax}
\providecommand{\BIBforeignlanguage}[2]{{%
\expandafter\ifx\csname l@#1\endcsname\relax
\typeout{** WARNING: IEEEtran.bst: No hyphenation pattern has been}%
\typeout{** loaded for the language `#1'. Using the pattern for}%
\typeout{** the default language instead.}%
\else
\language=\csname l@#1\endcsname
\fi
#2}}
\providecommand{\BIBdecl}{\relax}
\BIBdecl

\bibitem{fccspectrum}
\BIBentryALTinterwordspacing
{Federal Communications Commission}, \emph{Mobile Broadband: The Benefits of
  Additional Spectrum}, Oct, 2010. [Online]. Available:
  \url{https://apps.fcc.gov/edocs_public/attachmatch/DOC-302324A1.pdf}
\BIBentrySTDinterwordspacing

\bibitem{lifihaas}
H.~Haas, L.~Yin, Y.~Wang, and C.~Chen, ``What is {LiFi}?'' \emph{Journal of
  Lightwave Technology}, vol.~34, no.~6, pp. 1533--1544, March 2016.

\bibitem{tezcan5g}
T.~Cogalan and H.~Haas, ``Why would {5G} need optical wireless
  communications?'' in \emph{2017 IEEE 28th Annual International Symposium on
  PIMRC}, Montreal, Canada, Oct 2017, pp. 1--6.

\bibitem{lifiwg}
\BIBentryALTinterwordspacing
{The Institute of Electrical and Electronics Engineers, Inc.}, ``{802.11bb -
  Standard for Information Technology--Telecommunications and Information
  Exchange Between Systems Local and Metropolitan Area Networks--Specific
  Requirements - Part 11: Wireless LAN Medium Access Control (MAC) and Physical
  Layer (PHY) Specifications Amendment: Light Communications},'' accessed
  2018-07-24. [Online]. Available:
  \url{http://www.ieee802.org/15/pub/IEEE%20802_15%20WPAN%2015_7%20Revision1%20Task%20GroupOLD.htm}
\BIBentrySTDinterwordspacing

\bibitem{lifitig}
\BIBentryALTinterwordspacing
------, ``{IEEE 802.15 WPANTM 15.7 Revision: Short-Range Optical Wireless
  Communications Task Group (TG 7r1)},'' accessed 2018-04-11. [Online].
  Available:
  \url{http://www.ieee802.org/15/pub/IEEE%20802_15%20WPAN%2015_7%20Revision1%20Task%20GroupOLD.htm}
\BIBentrySTDinterwordspacing

\bibitem{infraredkahnbarry}
J.~M. Kahn and J.~R. Barry, ``Wireless infrared communications,''
  \emph{Proceedings of the IEEE}, vol.~85, no.~2, pp. 265--298, Feb 1997.

\bibitem{downlinkcheng}
C.~Chen, D.~A. Basnayaka, and H.~Haas, ``Downlink performance of optical
  attocell networks,'' \emph{Journal of Lightwave Technology}, vol.~34, no.~1,
  pp. 137--156, Jan 2016.

\bibitem{aveneauorientation}
C.~L. Bas, S.~Sahuguede, A.~Julien-Vergonjanne, A.~Behlouli, P.~Combeau, and
  L.~Aveneau, ``Impact of receiver orientation and position on visible light
  communication link performance,'' in \emph{2015 4th IWOW}, Istanbul, Turkey,
  Sept 2015, pp. 1--5.

\bibitem{orientationsoltani}
M.~D. Soltani, X.~Wu, M.~Safari, and H.~Haas, ``Access point selection in
  {Li-Fi} cellular networks with arbitrary receiver orientation,'' in
  \emph{2016 IEEE 27th Annual International Symposium on PIMRC}, Valencia,
  Spain, Sept 2016, pp. 1--6.

\bibitem{wangtilted}
J.~Y. Wang, Q.~L. Li, J.~X. Zhu, and Y.~Wang, ``Impact of receiver's tilted
  angle on channel capacity in {VLCs},'' \emph{Electronics Letters}, vol.~53,
  no.~6, pp. 421--423, 2017.

\bibitem{wangber}
J.~Y. Wang, J.~B. Wang, B.~Zhu, M.~Lin, Y.~Wu, Y.~Wang, and M.~Chen,
  ``Improvement of {BER} performance by tilting receiver plane for indoor
  visible light communications with input-dependent noise,'' in \emph{2017 IEEE
  ICC}, Paris, France, May 2017, pp. 1--6.

\bibitem{erogluorientation}
Y.~S. Eroglu, Y.~Yapici, and I.~G{\"{u}}ven{\c{c}}, ``Impact of random receiver
  orientation on visible light communications channel,'' \emph{CoRR}, vol.
  abs/1710.09764, 2017.

\bibitem{yapicinoma}
Y.~{Yapici} and I.~{Guvenc}, ``{Non-Orthogonal Multiple Access for Mobile {VLC}
  Networks with Random Receiver Orientation},'' \emph{ArXiv e-prints}, jan
  2018.

\bibitem{ardimaswcnc}
A.~A. Purwita, M.~D. Soltani, M.~Safari, and H.~Haas, ``Impact of terminal
  orientation on performance in {LiFi} systems,'' in \emph{2018 IEEE Wireless
  Communications and Networking Conference (WCNC)}, Barcelona, Spain, April
  2018, pp. 1--6.

\bibitem{MDSHandover}
M.~D. Soltani, H.~Kazemi, M.~Safari, and H.~Haas, ``Handover modeling for
  indoor {Li-Fi} cellular networks: The effects of receiver mobility and
  rotation,'' in \emph{2017 IEEE WCNC}, San Fransisco, USA, March 2017, pp.
  1--6.

\bibitem{ardimashandover}
A.~A. Purwita, M.~D. Soltani, M.~Safari, and H.~Haas, ``Handover probability of
  hybrid {LiFi/RF-based} networks with randomly-oriented devices,'' in
  \emph{2018 IEEE 87th VTC Spring (accepted)}, Porto, Portugal, June 2018, pp.
  1--5.

\bibitem{soltaniaccess}
M.~D. Soltani, A.~A. Purwita, I.~Tavakkolnia, H.~Haas, and M.~Safari, ``Impact
  of device orientation on error performance of {LiFi} systems,'' \emph{IEEE
  Access (to be submitted)}, 2018.

\bibitem{soltanitcom}
\BIBentryALTinterwordspacing
M.~D. Soltani, A.~A. Purwita, Z.~Zeng, H.~Haas, and M.~Safari, ``Modeling the
  random orientation of mobile devices: Measurement, analysis and {LiFi} use
  case,'' \emph{CoRR}, vol. abs/1805.07999, 2018. [Online]. Available:
  \url{http://arxiv.org/abs/1805.07999}
\BIBentrySTDinterwordspacing

\bibitem{orientationpeng}
B.~Peng and T.~K{\"u}rner, ``Three-dimensional angle of arrival estimation in
  dynamic indoor terahertz channels using a forward-backward algorithm,''
  \emph{IEEE Transaction on Vehicular Technology}, vol.~66, no.~5, pp.
  3798--3811, May 2017.

\bibitem{vaniceklssa}
P.~Van{\'i}{\v{c}}ek, ``Approximate spectral analysis by least-squares fit,''
  \emph{Astrophysics and Space Science}, vol.~4, no.~4, pp. 387--391, Aug 1969.

\bibitem{scargle1982}
J.~D. {Scargle}, ``{Studies in astronomical time series analysis. II -
  Statistical aspects of spectral analysis of unevenly spaced data},''
  \emph{Astrophysical Journal}, vol. 263, pp. 835--853, Dec. 1982.

\bibitem{craymerdissertation}
M.~R. Craymer, ``The least-squares spectrum, its inverse transform and
  autocorrelation function: Theory and some applications in geodesy,'' Ph.D.
  dissertation, University of Toronto, Canada, 1998.

\bibitem{baisch1999}
S.~Baisch and G.~H. Bokelmann, ``Spectral analysis with incomplete time series:
  an example from seismology,'' \emph{Computers \& Geosciences}, vol.~25,
  no.~7, pp. 739 -- 750, 1999.

\bibitem{vanderplas}
J.~T. {VanderPlas}, ``{Understanding the Lomb-Scargle Periodogram},''
  \emph{ArXiv e-prints}, Mar 2017.

\bibitem{hayesbook}
M.~H. Hayes, \emph{Statistical Digital Signal Processing and Modeling},
  1st~ed.\hskip 1em plus 0.5em minus 0.4em\relax New York, NY, USA: John Wiley
  \& Sons, Inc., 1996.

\bibitem{robertsclean}
D.~H. {Roberts}, J.~{Lehar}, and J.~W. {Dreher}, ``{Time Series Analysis with
  Clean - Part One - Derivation of a Spectrum},'' \emph{Astronomical Journal},
  vol.~93, p. 968, Apr 1987.

\bibitem{barrykahnsim}
J.~R. Barry, J.~M. Kahn, W.~J. Krause, E.~A. Lee, and D.~G. Messerschmitt,
  ``Simulation of multipath impulse response for indoor wireless optical
  channels,'' \emph{IEEE Journal on Selected Areas in Communications}, vol.~11,
  no.~3, pp. 367--379, Apr 1993.

\bibitem{highorderreflection}
Z.~Zhou, C.~Chen, and M.~Kavehrad, ``Impact analyses of high-order light
  reflections on indoor optical wireless channel model and calibration,''
  \emph{Journal of Lightwave Technology}, vol.~32, no.~10, pp. 2003--2011, May
  2014.

\bibitem{perinofdm}
J.~K. Perin, M.~Sharif, and J.~M. Kahn, ``Modulation schemes for single-laser
  100 {Gb/s} links: Multicarrier,'' \emph{Journal of Lightwave Technology},
  vol.~33, no.~24, pp. 5122--5132, Dec 2015.

\bibitem{predistortionelgala}
H.~Elgala, R.~Mesleh, and H.~Haas, ``Predistortion in optical wireless
  transmission using {OFDM},'' in \emph{2009 Ninth International Conference on
  Hybrid Intelligent Systems}, vol.~2, Shenyang, China, Aug 2009, pp. 184--189.

\bibitem{schulzefreqdomain}
H.~Schulze, ``Frequency-domain simulation of the indoor wireless optical
  communication channel,'' \emph{IEEE Transaction on Communications}, vol.~64,
  no.~6, pp. 2551--2562, June 2016.

\bibitem{carruthersiterative}
J.~B. Carruthers and P.~Kannan, ``Iterative site-based modeling for wireless
  infrared channels,'' \emph{IEEE Transaction on Antennas and Propagation},
  vol.~50, no.~5, pp. 759--765, May 2002.

\bibitem{laveneaubody}
C.~L. Bas, S.~Sahuguede, A.~Julien-Vergonjanne, A.~Behlouli, P.~Combeau, and
  L.~Aveneau, ``Human body impact on mobile visible light communication link,''
  in \emph{2016 10th International Symposium on CSNDSP}, Prague, Czech
  Republic, July 2016, pp. 1--6.

\bibitem{zhihongorientation}
Z.~Zeng, M.~D. Soltani, H.~Haas, and M.~Safari, ``Orientation model of mobile
  device for indoor {VLC} and millimetre wave systems,'' in \emph{2018 IEEE
  88th VTC Fall (accepted)}, Chicago, USA, 2018.

\bibitem{androidapp}
\BIBentryALTinterwordspacing
V.~Software, ``Physics toolbox sensor suite,'' accessed 2018-02-07. [Online].
  Available:
  \url{https://play.google.com/store/apps/details?id=com.chrystianvieyra.physicstoolboxsuite}
\BIBentrySTDinterwordspacing

\bibitem{eyer1999}
{Eyer, L.} and {Bartholdi, P.}, ``Variable stars: Which {Nyquist} frequency?''
  \emph{Astronomy and Astrophysics Supplement Series}, vol. 135, no.~1, pp.
  1--3, 1999.

\bibitem{babu2010}
P.~Babu and P.~Stoica, ``Spectral analysis of nonuniformly sampled data – a
  review,'' \emph{Digital Signal Processing}, vol.~20, no.~2, pp. 359 -- 378,
  2010.

\bibitem{gresty1990}
M.~Gresty and D.~Buckwell, ``Spectral analysis of tremor: understanding the
  results.'' \emph{Journal of Neurology, Neurosurgery \& Psychiatry}, vol.~53,
  no.~11, pp. 976--981, 1990.

\bibitem{dick2017}
O.~Dick, ``From healthy to pathology through a fall in dynamical complexity of
  involuntary oscillations of the human hand,'' \emph{Neurocomputing}, vol.
  243, pp. 142 -- 154, 2017.

\bibitem{hsictest}
K.~Chwialkowski and A.~Gretton, ``A kernel independence test for random
  processes,'' in \emph{Proceedings of the 31st International Conference on
  Machine Learning}, vol.~32, no.~2.\hskip 1em plus 0.5em minus 0.4em\relax
  Bejing, China: PMLR, 22--24 Jun 2014, pp. 1422--1430.

\bibitem{humandimension}
\BIBentryALTinterwordspacing
J.~Ilecko, ``The simulation of human gait in {Solid Works},'' accessed
  2018-02-07. [Online]. Available:
  \url{http://www.rusnauka.com/31_PRNT_2008/Tecnic/36223.doc.htm}
\BIBentrySTDinterwordspacing

\bibitem{cottonfabric}
\BIBentryALTinterwordspacing
R.~F. Kokaly, R.~N. Clark \emph{et~al.}, ``Cotton fabric {GDS437} white
  descript,'' accessed 2018-02-07. [Online]. Available:
  \url{https://crustal.usgs.gov/speclab/data/GIFplots/GIFplots_splib07a/ChapterA_ArtificialMaterials/splib07a_Cotton_Fabric_GDS437_White_ASDFRa_AREF.gif}
\BIBentrySTDinterwordspacing

\bibitem{humanskin}
D.~W.~A. Catherine C.~Cooksey, ``Reflectance measurements of human skin from
  the ultraviolet to the shortwave infrared (250 nm to 2500 nm),'' pp. 8734 --
  8734 -- 9, 2013.

\bibitem{propertieshair}
T.~van Kampen, ``Optical properties of hair,'' Master's thesis, Technische
  Universiteit Eindhoven, Netherland, 1997.

\bibitem{ledmodelflat}
J.~Vucic, C.~Kottke, S.~Nerreter, K.~D. Langer, and J.~W. Walewski, ``{513
  Mbit/s} visible light communications link based on {DMT}-modulation of a
  white {LED},'' \emph{Journal of Lightwave Technology}, vol.~28, no.~24, pp.
  3512--3518, Dec 2010.

\bibitem{ledmodelobrien}
H.~L. Minh, D.~O'Brien, G.~Faulkner, L.~Zeng, K.~Lee, D.~Jung, Y.~Oh, and E.~T.
  Won, ``{100-Mb/s NRZ} visible light communications using a postequalized
  white {LED},'' \emph{IEEE Photonics Technology Letters}, vol.~21, no.~15, pp.
  1063--1065, Aug 2009.

\bibitem{8gbpsislim}
M.~S. Islim, R.~X. Ferreira, X.~He, E.~Xie, S.~Videv, S.~Viola, S.~Watson,
  N.~Bamiedakis, R.~V. Penty, I.~H. White, A.~E. Kelly, E.~Gu, H.~Haas, and
  M.~D. Dawson, ``Towards 10{Gb/s} orthogonal frequency division
  multiplexing-based visible light communication using a {GaN} violet
  micro-{LED},'' \emph{Photon. Res.}, vol.~5, no.~2, pp. A35--A43, Apr 2017.

\bibitem{barrosofdm}
D.~J.~F. Barros, S.~K. Wilson, and J.~M. Kahn, ``Comparison of orthogonal
  frequency-division multiplexing and pulse-amplitude modulation in indoor
  optical wireless links,'' \emph{IEEE Transaction on Communications}, vol.~60,
  no.~1, pp. 153--163, January 2012.

\bibitem{erickcoverage}
L.~Yin and H.~Haas, ``Coverage analysis of multiuser visible light
  communication networks,'' \emph{IEEE Transaction on Wireless Communications},
  vol.~PP, no.~99, pp. 1--1, 2017.

\bibitem{ardimasglobecom}
A.~A. Purwita, C.~Chen, D.~A. Basnayaka, and H.~Haas, ``Aggregate signal
  interference of downlink {LiFi} networks,'' in \emph{GLOBECOM 2017 - 2017
  IEEE Global Communications Conference}, Singapore, Singapore, Dec 2017, pp.
  1--6.

\bibitem{ericknoma}
L.~Yin, W.~O. Popoola, X.~Wu, and H.~Haas, ``Performance evaluation of
  non-orthogonal multiple access in visible light communication,'' \emph{IEEE
  Transaction on Communications}, vol.~64, no.~12, pp. 5162--5175, Dec 2016.

\bibitem{aveneaubody}
C.~L. Bas, S.~Sahuguede, A.~Julien-Vergonjanne, A.~Behlouli, P.~Combeau, and
  L.~Aveneau, ``Human body impact on mobile visible light communication link,''
  in \emph{2016 10th International Symposium on CSNDSP}, Prague, Czech
  Republic, July 2016, pp. 1--6.

\end{thebibliography}

\end{document}